\documentclass{revtex4}

\usepackage{graphicx}
\def\thm{\theta^-}
\def\thp{\theta^+}
\def\cp{c_+}
\def\sp{s_+}
\def\sm{s_-}
\def\cm{c_-}
\def\php{P_H^+}
\def\phm{P_H^-}
\def\pvp{P_V^+}
\def\pvm{P_V^-}

\def\gam{\gamma}

\def\bit{\begin{itemize}}
\def\eit{\end{itemize}}
\def\zstar{Z^\star}
\def\wstar{W^\star}
\def\cnone{\chi^0}

\def\what{\widehat}
\def\fbi{~{\rm fb}^{-1}}
\def\abi{~{\rm ab}^{-1}}
\def\call{{\cal L}}
\def\lyear{L_{\rm year}}
\def\sig{\sigma}
\def\gamhsmtot{\Gamma_{\hsm}^{\rm tot}}
\def\br{{\rm BF}}
\def\gamhtot{\Gamma_{\h}^{\rm tot}}
\def\gamhltot{\Gamma_{\hl}^{\rm tot}}
\def\hl{h^0}
\def\ha{A^0}
\def\hh{H^0}
\def\hpm{H^{\pm}}
\def\mhl{m_{\hl}}
\def\mha{m_{\ha}}
\def\mhh{m_{\hh}}

\def\h{h}
\def\mh{m_{\h}}
\def\epem{e^+e^-}
\def\mupmum{\mu^+\mu^-}
\def\anti{\overline}
\def\tanb{\tan\beta}
\def\what{\widehat}
\def\rts{\sqrt s}
\def\hsm{h_{SM}}
\def\mhsm{m_{\hsm}}
\def\mev{~{\rm MeV}}
\def\gev{~{\rm GeV}}
\def\tev{~{\rm TeV}}

\def\gtap{\raisebox{-.4ex}{\rlap{$\sim$}} \raisebox{.4ex}{$>$}}
\def\mumu{$\mu^+\mu^-$}
\def\srts{\sigma_{\!\!\!\sqrt s}^{\vphantom y}}
\def\lsim{\alt}
\def\gsim{\agt}
\def\beq{\begin{equation}}
\def\eeq{\end{equation}}
\def\bea{\begin{eqnarray}}
\def\eea{\end{eqnarray}}
\def\mumu{\mu^-\mu^+}
\def\tautau{\tau^-\tau^+}
\begin{document}
\bibliographystyle{revtex}


\title{Physics of Higgs Factories}


\author{V. Barger$^1$, M. S. Berger$^2$, J. F. Gunion$^3$, and T. Han$^1$}
\affiliation{$^1$Physics Department, University of Wisconsin, 
Madison, WI 53706\\
$^2$Physics Department, Indiana University, 
Bloomington, IN 47405\\
$^3$Physics Department, University of California, 
Davis, CA 95616}

\title{Physics of Higgs Factories\footnote{Submitted to the Proceedings
of ``The Future of Particle Physics'', Snowmass 2001, E1 group.}\thanks{This work was supported in part by the U.S. Department of Energy.}
}


\begin{abstract}
We outline the unique role of a muon collider
as a Higgs factory for Higgs boson resonance production in the $s$-channel.
Physics examples include: the precision measurements of the Higgs
mass and total width, and the resulting ability to discriminate between
the SM-like Higgs bosons of
different models such as between a light SM Higgs boson and the 
light Higgs boson of the MSSM; the determination of the
spin and coupling via the $h\to \tautau$ decay mode; differentiation of two
nearly degenerate heavy Higgs bosons by an energy scan; and 
the ability to explore a general extended Higgs
sector, possibly with CP-violating couplings. 
The muon collider Higgs factory could perform measurements
that would be highly complementary to
Higgs studies at the LHC and LC; it would be likely to 
play a very crucial role in fully understanding the Higgs sector.
\end{abstract}

\hbox to \hsize{
%
%
\hfill
$\vcenter{\normalsize
\hbox{\bf UCD-2001-10} 
\hbox{\bf hep-ph/0110340}
\hbox{October, 2001}
}$
}

\maketitle

\section{Introduction}
%
%

%
%

A muon collider with c.m. energy centered at the Higgs
boson mass offers a unique opportunity to produce Higgs bosons in the 
$s$-channel and thereby measure the Higgs masses, total width and several 
partial widths to very high precision. In the event 
that only a SM-like Higgs boson is discovered and its properties measured
at the Tevatron, the LHC, and a LC, it may 
prove essential to build a muon 
collider to fully explore the Higgs sector.
In particular, the very narrow width of a Standard 
Model (SM) Higgs bosons cannot be measured {\it directly} at the Large Hadron 
Collider (LHC) or at a future Linear Collider (LC). 
Furthermore, there are regions of 
parameter space for which it will be impossible for either the LHC or a 
LC to discover the heavier Higgs bosons of supersymmetry or,
in the case of a general two-Higgs-doublet or more extended model,
Higgs bosons of any mass with small or zero $VV$ coupling. 

The value of a future Higgs factory should be 
discussed in light of recent experimental 
data. While by no means definitive,
recent experimental results point in promising directions for 
Higgs factories. First, there is the $\gsim 2\sigma$
statistical evidence from 
LEP\cite{Barate:2000ts,Abreu:2001fw,Acciarri:2000ke,Abbiendi:2001ac,Okpara:2001jf}  
for a Higgs boson near $m_H\simeq 115$~GeV. Such a mass
is in the optimal range for study at a Higgs factory
and it is for such a low mass that the muon collider factory
option would add the most information to data from the LHC and a LC. 
First, $115\gev$ is sufficiently above the $Z$-pole that the background from 
$Z$ production and decay to $b\overline{b}$ is not so 
large, and the mass is sufficiently below the $WW^\star$ threshold  that 
the decay width remains small and the ability of the muon collider to
achieve a very narrow beam energy spread can be exploited.
Second, it is for masses below $120\gev$ that the LC will have 
difficulty getting a precision measurement of the Higgs to $WW^*$
branching ratio, resulting in
large error for the indirect determination of the total Higgs width.
Of course, a Higgs boson in this mass range, and having substantial 
$VV$ coupling, is also the most natural interpretation of
current precision electroweak data. On the theoretical side,  
a Higgs mass of $\sim 115$~GeV is very suggestive
of supersymmetry. In the Minimal Supersymmetric Model (MSSM) such a 
mass is near
the theoretical upper limit of $m_H<130$~GeV, and would indicate
a value of the supersymmetry parameter $\tan \beta$
substantially above 1 (assuming stop masses $\lsim 1\tev$). 

A Higgs with mass $\sim 115$~GeV in the context of  
a large-$\tanb$ supersymmetry scenario
would mesh nicely with recent evidence for an anomalous magnetic moment of the 
muon\cite{Brown:2001mg}  
that deviates from the Standard Model prediction. The $2.6\sigma$
discrepancy is naturally accounted for provided $\tan \beta$ is 
relatively large (and superparticle masses are not too heavy).
More specifically, a 
supersymmetric interpretation of this discrepancy with the 
SM prediction implies the following relationship between the mass scale 
$\tilde{m}$ of supersymmetric particles contributing to the one-loop anomalous
magnetic moment diagram and $\tan \beta$\cite{Czarnecki:2001pv},
\begin{eqnarray}
&&\tan \beta \left ({{100~{\rm GeV}}\over {\tilde{m}}}\right )^2=3.3\pm 1.3
\nonumber
\end{eqnarray}
Furthermore, if the anomalous magnetic moment is explained by supersymmetry
the value of the Higgs mass parameter $\mu $ of supersymmetric models has a
sign which is consistent with the constraints from the radiative decays, 
$b\to s \gamma$.
Thus, a consistent picture begins to emerge 
suggesting low-energy supersymmetry with a Higgs boson in the predicted
mass range.

While these recent experimental data are not definitive, they do point to 
an interesting scenario whereby a muon collider might prove essential to the 
understanding of the Higgs sector of a supersymmetric model. The muon collider
could perform at least two measurements crucial for
detailing a SUSY Higgs sector: 
(1) accurately measuring the properties of a light 
SM-like Higgs boson and distinguishing it from a supersymmetric Higgs bosons,
and (2) discovering heavy Higgs bosons of supersymmetry and accurately
measuring their properties. 

\section{Muon Colliders}
Muon colliders have a number of unique features that make them attractive
candidates for future accelerators\cite{Ankenbrandt:1999as}. 
The most important and fundamental of 
these derive from the large mass of the 
muon in comparison to that of the electron.
This leads to: a) the possibility of extremely narrow beam
energy spreads, especially at beam energies below $100\gev$;
b) the possibility of accelerators 
with very high energy; c) the possibility of employing
storage rings at high energy; d) the possibility of using decays of accelerated
muons to provide a high luminosity source of neutrinos (under
active consideration as reviewed elsewhere); e) increased 
potential for probing physics in which couplings increase with
mass (as does the SM $\hsm f\anti f$ coupling).

\begin{figure*}[hbt!]
\begin{center}
\includegraphics[width=3.5in]{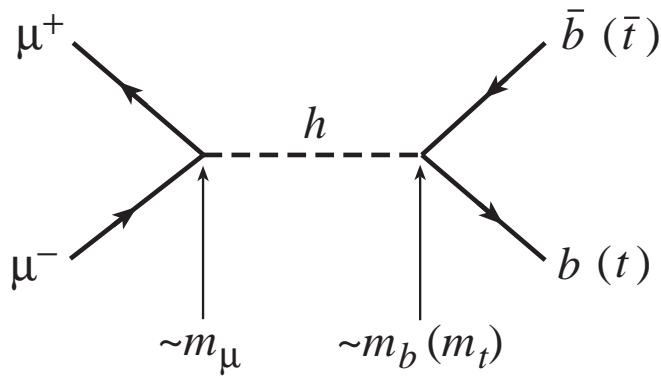}
\caption{\label{feyn-diag}Feynman diagram for $s$-channel production of a Higgs boson.}
\end{center}
\end{figure*}

Here our focus is on the Higgs sector.
The relatively large mass of the muon compared to the mass of the electron
means that the coupling of Higgs bosons to $\mu^+\mu^-$ is very 
much larger than to $e^+e^-$, implying much larger $s$-channel Higgs
production rates at a muon collider as compared to an electron collider
[see Fig.~\ref{feyn-diag}].
For Higgs bosons with a very small MeV scale width, 
such as a light SM Higgs boson,
production rates in the $s$-channel 
are further enhanced by the 
muon collider's ability to achieve beam energy spreads
comparable to the tiny Higgs width. 
In addition, there is little bremsstrahlung, 
and the beam energy can be tuned to one part
in a million through continuous spin-rotation measurements\cite{Raja:1998ip}.
Due to these important qualitative differences
between the two types of machines, only muon colliders can be 
advocated as potential $s$-channel
Higgs factories capable of determining the mass and decay width
of a Higgs boson to very high precision\cite{Barger:1997jm,Barger:1995hr}.
High rates of Higgs production at $\epem$ colliders rely on
substantial $VV$Higgs coupling for the
$Z+$Higgs (Higgs-strahlung) or $WW\to$Higgs ($WW$ fusion) reactions,
In contrast, a $\mupmum$ collider can provide a factory for producing
a Higgs boson with little or no $VV$ coupling so long as it
has SM-like (or enhanced) $\mupmum$ couplings.

Of course, there is a trade-off between small beam energy spread,
$\delta E/E=R$, and luminosity. Current estimates for yearly
integrated luminosities (using
$\call=1\times 10^{32}$cm$^{-2}$s$^{-1}$ as implying $ L=1\fbi/{\rm yr}$) are:
$\lyear\gsim 0.1,0.22,1 \fbi$ at $\rts\sim 100\gev$
for beam energy resolutions of $R=0.003\%,0.01\%,0.1\%$, respectively;
$\lyear\sim 2,6,10 \fbi$ at $\rts\sim 200,350,400\gev$, respectively, for 
$R\sim 0.1\%$.
Despite this, studies show that for small Higgs width the $s$-channel
production rate (and statistical significance over background) is maximized
by choosing $R$ to be such that $\srts\lsim \gamhtot$. In particular,
in the SM context this corresponds to $R\sim 0.003\%$ for $\mhsm\lsim 120\gev$.

If the $\mh\sim 115\gev$ LEP signal is real or if the 
interpretation of the precision
electroweak data as an indication of a light Higgs boson (with
substantial $VV$ coupling) is valid,~\footnote{Even
in a two-doublet extension of the minimal one-doublet SM Higgs
sector, parameters can be chosen so that the only
light Higgs boson has no $VV$ coupling and yet good agreement
with precision electroweak data maintained \cite{Chankowski:2000an}.}
then both $\epem$ and $\mupmum$ colliders will be valuable.
In this scenario the Higgs boson would have been discovered at a previous 
higher energy collider (possibly a muon collider
running at high energy), and then the Higgs factory
would be built with a center-of-mass energy 
precisely tuned to the Higgs boson mass.\footnote{If the higher energy
muon collider has already been constructed, this would simply require
construction of a small storage ring tuned to the appropriate energy.}
The most likely scenario is that the Higgs boson 
is discovered at the LHC via gluon fusion
($gg\to H$) or perhaps 
earlier at the Tevatron via associated production 
($q\bar{q}\to WH, t\overline{t}H$), and its mass is determined to an 
accuracy of about 100~MeV. If a linear collider has also observed the Higgs
via the Higgs-strahlung process ($e^+e^-\to ZH$), one might know the Higgs 
boson mass to better than 50~MeV with an integrated luminosity of 
$500$~fb$^{-1}$.
The muon collider would be optimized to run at $\sqrt{s}\approx m_H$, and this
center-of-mass energy would be varied over a narrow range
so as to scan over the Higgs resonance (see Fig.~\ref{mhsmscan} below). 

\section{Higgs Production}

The production of a Higgs boson (generically denoted $\h$)
in the $s$-channel with interesting rates is  
a unique feature of a muon collider \cite{Barger:1997jm,Barger:1995hr}. 
The resonance cross section is
\begin{equation}
\sigma_h(\sqrt s) = {4\pi \Gamma(h\to\mu\bar\mu) \, \Gamma(h\to X)\over
\left( s - m_h^2\right)^2 + m_h^2 \left(\Gamma_{\rm tot}^h \right)^2}\,.
\label{rawsigform}
\end{equation}
In practice, however, there is a Gaussian spread ($\srts$) to
the center-of-mass energy and one must compute the
effective $s$-channel Higgs cross section after convolution 
assuming some given central value of $\rts$:
\bea
\anti\sigma_h(\sqrt s) & =& {1\over \sqrt{2\pi}\,\srts} \; \int \sigma_h  
(\sqrt{\what s}) \; \exp\left[ -\left( \sqrt{\what s} - \sqrt s\right)^2 \over  
2\sigma_{\sqrt s}^2 \right] d \sqrt{\what s}~~~
\stackrel{\rts=\mh}{\simeq}  ~~~ {4\pi\over m_h^2} \; {\br(h\to\mu\bar\mu) \,
\br(h\to X) \over \left[ 1 + {8\over\pi} \left(\srts\over\gamhtot 
\right)^2 \right]^{1/2}} \,.
\label{sigform}
\eea
\begin{figure*}[h!]
\begin{center}
\includegraphics[width=3.5in]{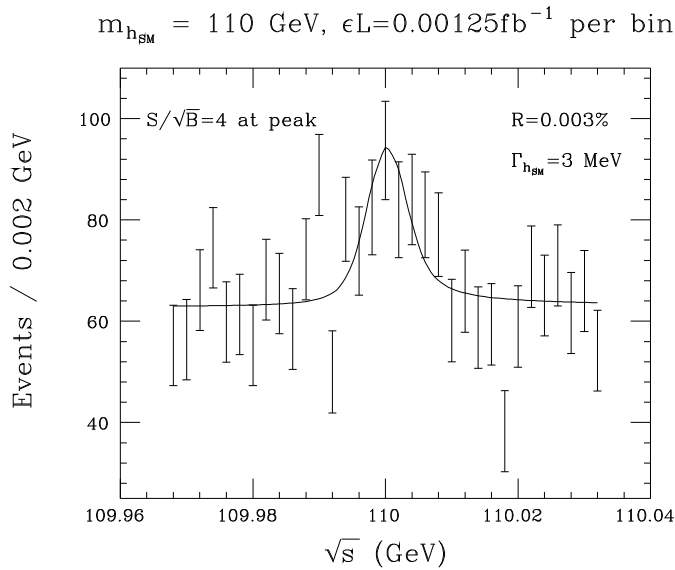}
\caption{
Number of events and statistical errors in the $b\overline{b}$
final state as a function
of $\protect\rts$ in the vicinity of $\mhsm=110\gev$,
assuming $R=0.003\%$,
and $\epsilon L=0.00125$~fb$^{-1}$ at each data point.
\label{mhsmscan}}
\end{center}
\end{figure*}
It is convenient to express $\srts$ in 
terms of the root-mean-square (rms) Gaussian spread
of the energy of an individual beam, $R$: 
\begin{equation}
\srts = (2{\rm~MeV}) \left( R\over 0.003\%\right) \left(\sqrt s\over  
100\rm~GeV\right) \,.
\end{equation}
From Eq.~(\ref{rawsigform}), it is apparent that a
resolution $\srts \lsim \gamhtot$ is needed to be
sensitive to the Higgs width. Further, Eq.~(\ref{sigform}) implies that
$\anti\sigma_h\propto 1/\srts$ for $\srts>\gamhtot$ {\it and}
that large event rates are only possible if $\gamhtot$ is not so large
that $\br(\h\to \mu\anti\mu)$ is extremely suppressed.
The width of a light SM-like Higgs is very small (e.g. a few MeV
for $\mhsm\sim 110\gev$), implying the need for $R$
values as small as $\sim 0.003\%$ for studying a light SM-like $\h$.
Fig.~\ref{mhsmscan} illustrates the result for the SM Higgs boson 
of an initial centering scan over $\rts$ values
in the vicinity of $\mhsm=110\gev$.
This figure dramatizes: a)  that the beam energy spread must be very small
because of the very small $\gamhsmtot$ (when $\mhsm$ is small
enough that the $WW^\star$ decay
mode is highly suppressed); b) that we require
the very accurate {\it in situ} determination 
of the beam energy to one part in a million through the spin 
precession of the muon noted earlier in order to perform the scan
and then center on $\rts=\mhsm$ with a high degree of stability.

If the $\h$ has SM-like couplings to $WW$, its width will
grow rapidly for $\mh>2m_W$ and its $s$-channel production cross
section will be severely suppressed by the resulting 
decrease of $\br(\h\to\mu\mu)$. 
More generally, any $\h$ with SM-like or larger $\h\mu\mu$ coupling
will retain a large $s$-channel production rate when 
$\mh>2m_W$ only if the $\h WW$ coupling becomes 
strongly suppressed relative to the $\hsm WW$ coupling.

The general theoretical prediction within supersymmetric models is that the 
lightest supersymmetric Higgs boson $\hl$ will
be very similar to the $\hsm$ when the other Higgs bosons are
heavy.  This `decoupling limit' is very likely to arise if the
masses of the supersymmetric particles are large (since the Higgs
masses and the superparticle masses are typically similar in
size for most boundary condition choices).
Thus, $\hl$ rates will be very similar to $\hsm$ rates.
In contrast, the heavier Higgs bosons in a typical supersymmetric model
decouple from $VV$ at large mass  and remain reasonably
narrow. As a result, their $s$-channel production rates remain large.

\begin{figure*}[hbt!]
\begin{center}
\includegraphics[width=5in]{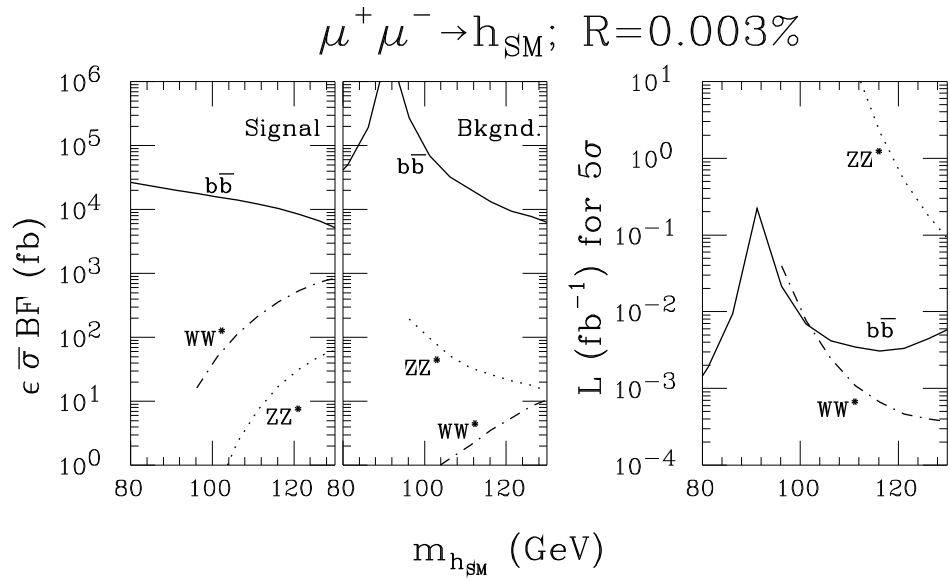}
\caption[The SM Higgs cross sections and backgrounds in $b\bar b,\ WW^*$  
and $ZZ^*$. ]{The SM Higgs cross sections and backgrounds in $b\bar b,\ WW^*$  
and $ZZ^*$. Also shown is the luminosity needed for a 5~standard deviation  
detection in $b\bar b$. From Ref.~\cite{Barger:1997jm}. 
\label{sm-higgs}}
\end{center}
\end{figure*}

For a SM-like $\h$, at $\sqrt s = \mh \approx 115$~GeV
and $R=0.003\%$, the $b\bar b$ final state rates are
\vspace{-.05in}
\begin{eqnarray}
\rm signal &\approx& 10^4\rm\ events\times L(fb^{-1})\,,\\
\rm background &\approx& 10^4\rm\ events\times L(fb^{-1})\,,
\end{eqnarray}
The SM Higgs cross sections and backgrounds are shown 
in Fig.~\ref{sm-higgs} for $R=0.003\%$ and
$\mhsm$ values such that the dominant decay mode is $b\overline{b}$.

\section{The Muon Collider Role}
An assessment of the need for a Higgs factory requires that one detail the 
unique capabilities of a muon collider versus the other possible future 
accelerators as well as comparing the abilities of all the machines to 
measure the same Higgs properties. 
Muon colliders and a Higgs factory in particular
would only become operational after the LHC physics program is well-developed 
and quite possibly after a linear collider program is mature as well. So one
important question is the following: if
a SM-like Higgs boson and, possibly, important
physics beyond the Standard Model have been discovered at the LHC and perhaps 
studied at a linear collider, what new information could a Higgs factory 
provide?

The $s$-channel production process allows one to determine the mass, 
total width, and the cross sections
$\overline \sig_h(\mupmum\to\h\to X)$ 
for several final states $X$ 
to very high precision. The Higgs mass, total width and the cross sections 
can be used to constrain the parameters of the Higgs sector. 
For example, in the MSSM their precise values will
constrain the Higgs sector parameters
$\mha$ and $\tanb$ (where $\tanb$ is 
the ratio of the two vacuum expectation values (vevs) of the 
two Higgs doublets of the MSSM). The main question is whether these
constraints will be a valuable addition to LHC and LC constraints.

The expectations for the luminosity available at linear colliders has risen 
steadily. The most recent studies assume an integrated luminosity of some
$500$~fb$^{-1}$ corresponding to 1-2 years of running at a 
few$\times100$~fb$^{-1}$ 
per year. This luminosity results in the production of greater than $10^4$
Higgs bosons per year through the Bjorken Higgs-strahlung process, 
$e^+e^-\to Z\h$, provided the Higgs boson is kinematically accessible. This is 
comparable or even better than can be achieved with the current machine
parameters for a muon collider operating at the Higgs resonance; in fact, 
recent studies have described high-luminosity linear colliders as ``Higgs
factories,'' though for the purposes of this report, we will reserve this term
for muon colliders operating at the $s$-channel Higgs resonance. 

A linear collider with such high luminosity can certainly perform quite 
accurate measurements of certain Higgs parameters such as the Higgs mass, 
couplings to gauge bosons, couplings to heavy quarks, 
etc.\cite{Battaglia:2000jb}.
Precise measurements of the couplings of the Higgs boson to the Standard 
Model particles is an important test of the mass generation mechanism.
In the Standard Model with one Higgs doublet, this coupling is proportional 
to the particle mass. In the more general case there can be mixing angles
present in the couplings. Precision measurements of the couplings can 
distinguish the Standard Model Higgs boson from the SM-like
Higgs boson typically present in a more general model. If 
deviations are found, their magnitude can be extremely crucial
for constraining the parameters of the more general Higgs sector. In
particular, it might be possible to estimate the masses of the
other Higgs bosons of the extended Higgs sector, thereby allowing
a more focused search for them.

\begin{table*}[h!]
\begin{center}
\caption[]{Achievable relative
uncertainties for a SM-like $\mh=110$~GeV for measuring the
Higgs boson mass and total width
for the LHC, LC (500~fb$^{-1}$), and the muon collider (0.2~fb$^{-1}$). 
}\label{unc-table}
\protect\protect
\begin{tabular}{cccc}
\hline
\ & LHC & LC & $\mu^+\mu^-$\\
$\mh$ & $9\times 10^{-4}$ & $3\times 10^{-4}$ & $1-3\times 10^{-6}$ \\
$\gamhtot$ & $>0.3$ & 0.17 & 0.2 \\
\hline
\end{tabular}
\end{center}
\end{table*}

The accuracies possible at different colliders
for measuring $\mh$ and $\gamhtot$ of
a SM-like $\h$ with $\mh\sim 110\gev$ are given in Table~\ref{unc-table}.
To achieve these accuracies, one first determines the Higgs mass 
to about 1~MeV by the preliminary scan
illustrated in Fig.~\ref{mhsmscan}. Then, a dedicated
three-point fine scan\cite{Barger:1997jm} near the resonance peak
using $L\sim 0.2\fbi$
of integrated luminosity (corresponding to a few years of operation)
would be performed.
For a SM Higgs boson with a mass sufficiently below the $WW^\star$ 
threshold, the Higgs total width is very small (of order several MeV), and the 
only process where it can be measured {\it directly} is in the $s$-channel
at a muon collider. Indirect determinations at the LC can have
higher accuracy once $\mh$ is large enough that the $WW^\star$ mode
rates can be accurately measured, requiring $\mh>120\gev$.
This is because at the LC the total width must be determined 
indirectly by measuring a partial width and a branching fraction, and then 
computing the total width,
\begin{eqnarray}
&&\Gamma _{tot}={{\Gamma(h\to X)}\over {BR(h\to X)}}\;,
\end{eqnarray} 
for some final state $X$. For a Higgs boson so light that the $WW^\star$ decay
mode is not useful, then the total width measurement would probably require
use of the $\h\to\gamma \gamma $ decays\cite{Gunion:1996cn}. This would require
information from a photon collider as well as the LC
and a small error is not possible. For $\mh\lsim 115\gev$,
the muon collider can measure the total width of the Higgs boson 
with greater precision than can be achieved using the indirect $\gam\gam$ mode
technique at the LC, and would be
a very valuable input for precision tests of the Higgs sector.
In particular, since all the couplings 
of the Standard Model $\hsm$ are known, $\gamhsmtot$
is precisely predicted. Therefore, the precise determination of 
$\gamhtot$ obtained by this scan
would be an important test of the Standard Model, and any deviation
would be evidence for a nonstandard Higgs sector (or other new physics).

In fact, a muon collider of limited luminosity
can remain more than competitive with LHC + LC 
for discriminating between the SM $\hsm$ and some SM-like $\h$
even for $\mh$ values such that the LC obtains a good measurement
of $WW^\star$ rates. 
As it happens, for $X=b\anti b$ there is a fortuitous compensation
that results in $\overline \sig_h(\mupmum\to\h\to b\anti b)$  
being almost completely
independent of the somewhat uncertain $b$ quark mass. Very roughly,
larger $m_b$ means larger $\br(\h\to b\anti b)$ but also larger $\gamhtot$.
The latter implies a smaller 
convoluted cross section $\overline\sig_h(\mupmum\to\h)$ (i.e. 
before including the branching ratio). Further, larger $\gamhtot$
means less damping because of beam energy spread.  The result is
that $\overline \sig_h(\mupmum\to\h\to b\anti b)$ is essentially
independent of the input $m_b$ value (within reasonable limits)
\cite{Autin:1999ci}. 
\begin{figure*}[h!]
\begin{center}
\includegraphics[width=3.0in]{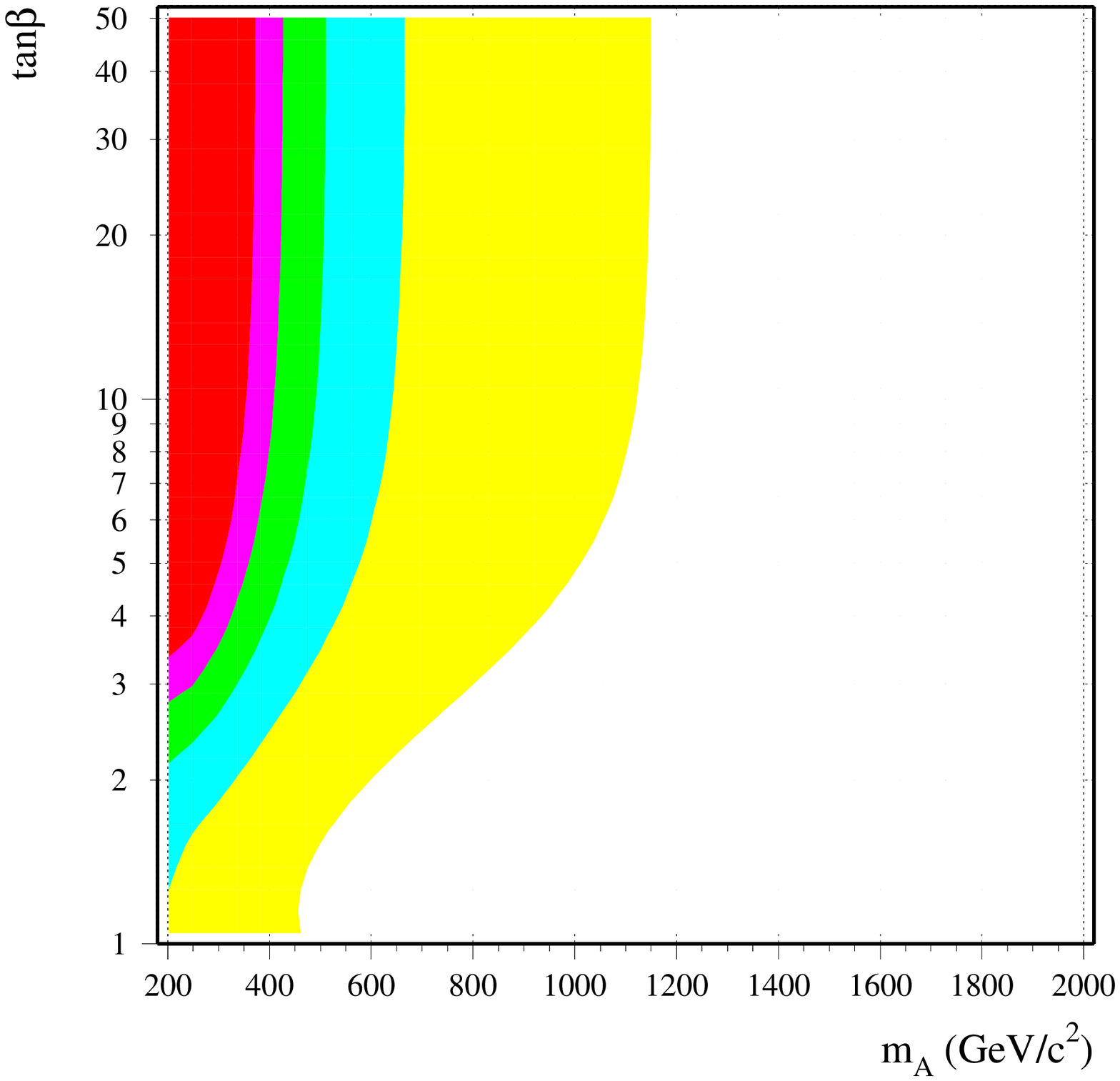}
\includegraphics[width=3.0in]{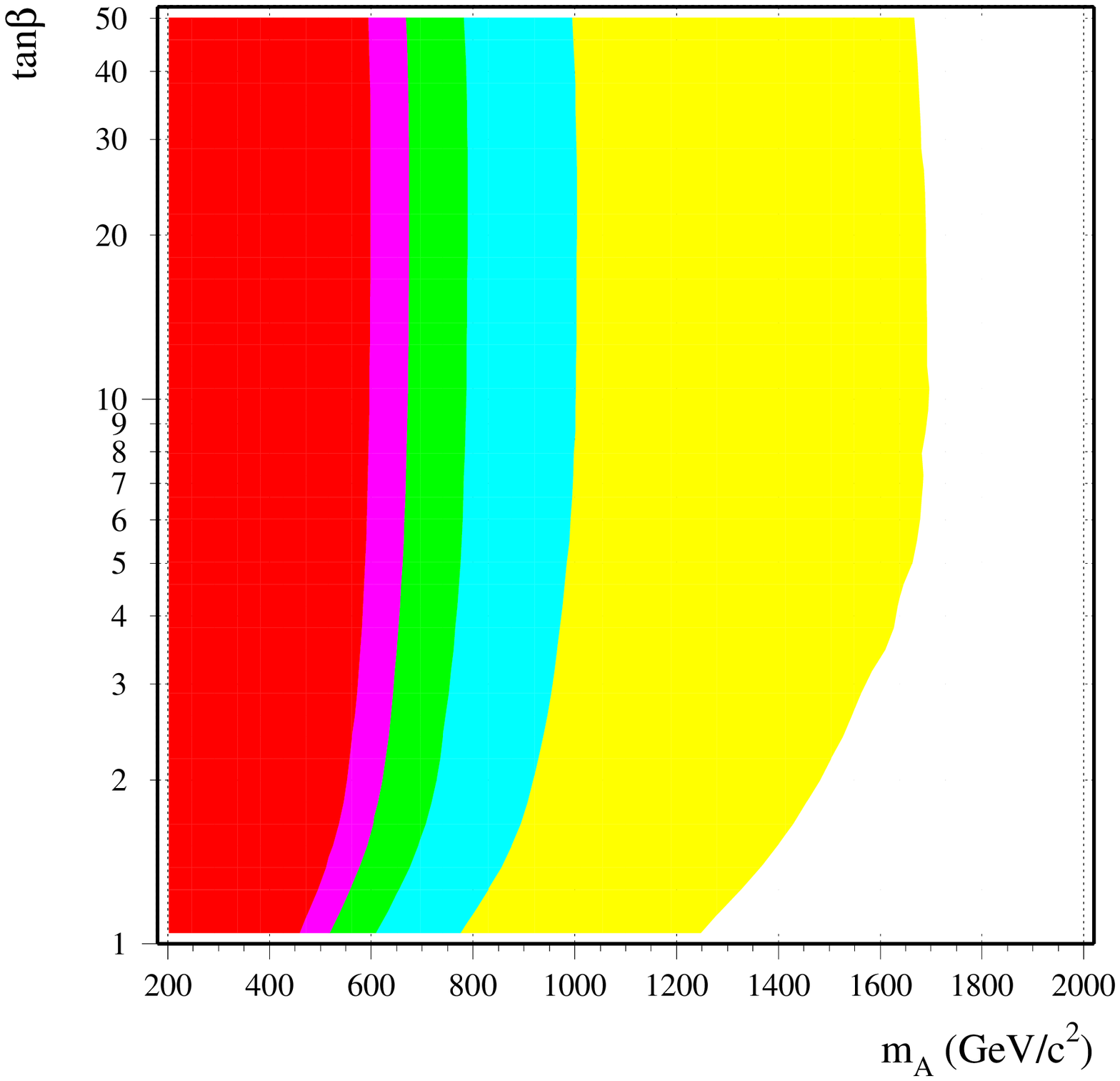}
\vspace*{.02in}
\includegraphics[width=3.0in]{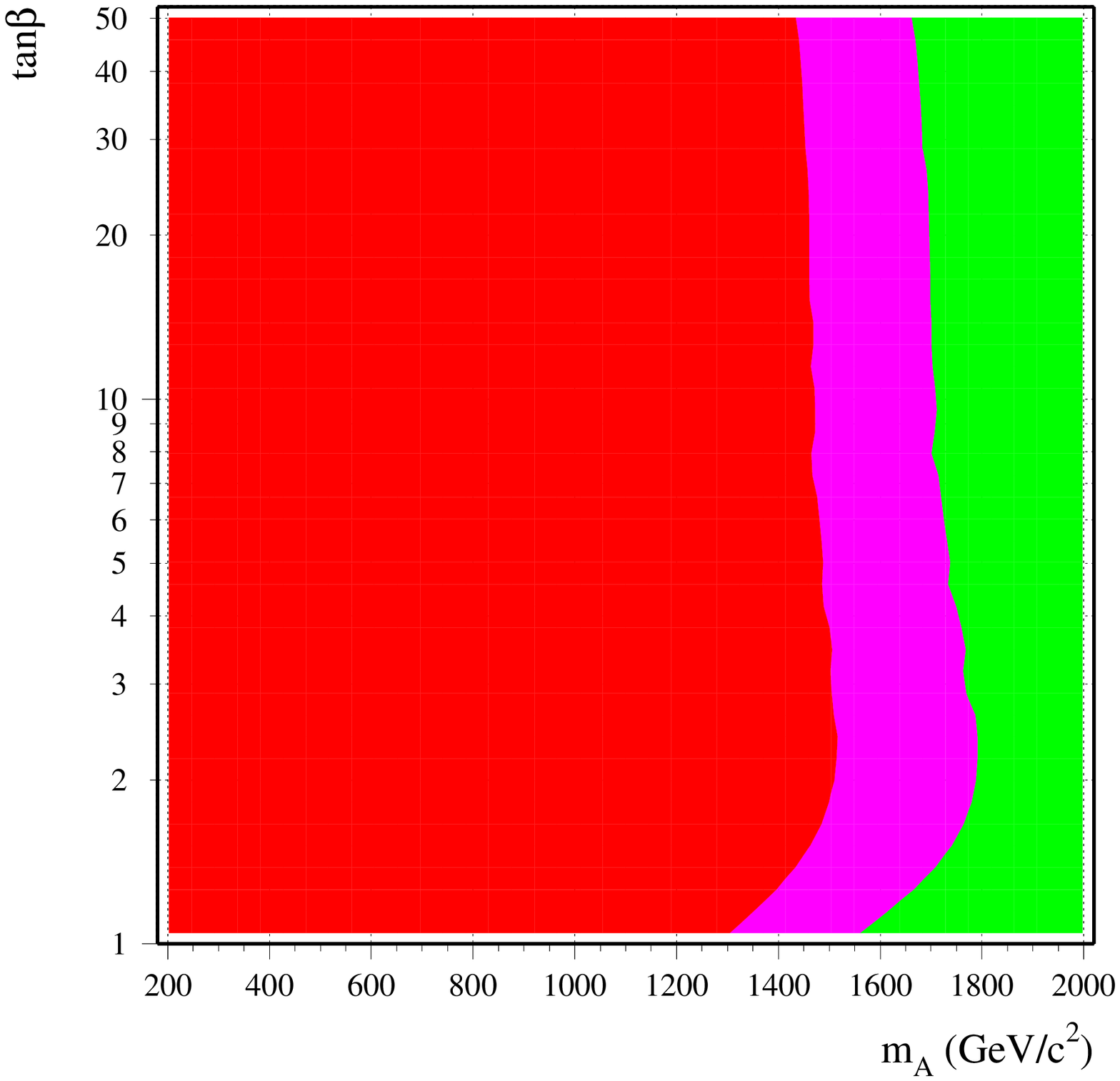}
\caption[The $\mha-\tanb$ discrimination]
{The $\mha-\tanb$ discrimination from the measurements at 
(a) LHC($300\fbi$)+LC($500\fbi$), 
(b) 0.2~fb$^{-1}$ at a muon collider, and (c) 10~fb$^{-1}$ at 
a muon collider. The exclusion regions (starting from the left) are
$>5\sigma$, $4-5\sigma$, $3-4\sigma$, $2-3\sigma$, and $1-2\sigma$.
From Ref.~\cite{Autin:1999ci}.
\label{bound}}
\end{center}
\end{figure*}
As a result, 
the precise measurement of $\overline \sig_h(\mupmum\to\h\to b\anti b)$
at a muon collider might provide the best single discriminator between
the SM Higgs and a SM-like Higgs.  This is nicely illustrated in the context
of the MSSM. For a Higgs mass of 110~GeV,
and assuming a typical soft-supersymmetry-breaking scenario,
Fig.~\ref{bound} shows the resulting excluded regions 
of $\mha$ for the (a) LHC+LC, 
(b) with a muon collider with 0.2~fb$^{-1}$ integrated luminosity, and 
(c) with a muon collider with 10~fb$^{-1}$ integrated luminosity.

Some comments on these results are appropriate. First,
one should note that the measurement
of $\gamhtot$ ($\pm 0.5$~MeV, i.e $\pm 20\%$) 
at the muon collider is not nearly so
powerful a discriminator as the $\pm 3.5\%$ ($\pm 0.5\%$) measurement
of $\overline \sig_h(\mupmum\to\h\to b\anti b)$ at $L=0.2\fbi$ ($10\fbi$).
Second, as $\mh$ increases and the $WW^\star$
decay mode becomes more prominent, much more accurate determinations
of partial width ratios and the total width become possible at LHC+LC
and the LHC+LC exclusion regions move rapidly to higher $\mha$,
but at best becoming comparable to the $0.2\fbi$ muon collider exclusion
regions. Third, the conclusion
that with higher luminosities than the $0.1\fbi$ per year
currently envisioned for the Higgs factory this 
discriminator would have incredible sensitivity to $\mha$ assumes
that systematic errors for
the absolute cross section will be smaller than the statistical errors.
Fourth, we should note that
there are high $\tanb$ scenarios in which decoupling sets
in very early in $\mha$ and no machine would be able to set a lower
bound on $\mha$; in particular, for such scenarios it would
be incorrect to conclude that the absence of deviations 
with respect to SM expectations implies
that $\mha\sim\mhh$ would be such that $\mha+\mhh>\rts$ so that
$\epem\to\hh\ha$ pair production is forbidden at a $\rts=500\gev$ LC. 
Finally, if there was a very light neutralino such that $\hl\to \cnone\cnone$
decays were possible, this would be known ahead of time and the
$\mupmum\to\hl\to b\anti b$ rate prediction within the SUSY context
would have to be corrected to
very high precision to account for these additional decays.
SUSY loop corrections to the $b\anti b$ coupling might also have
to be accounted for to high precision if the SUSY spectrum turns
out to be light.  But these last two caveats also apply to
the LC measurements.

\begin{figure*}[h!]
\begin{center}
\includegraphics[width=2.65in]{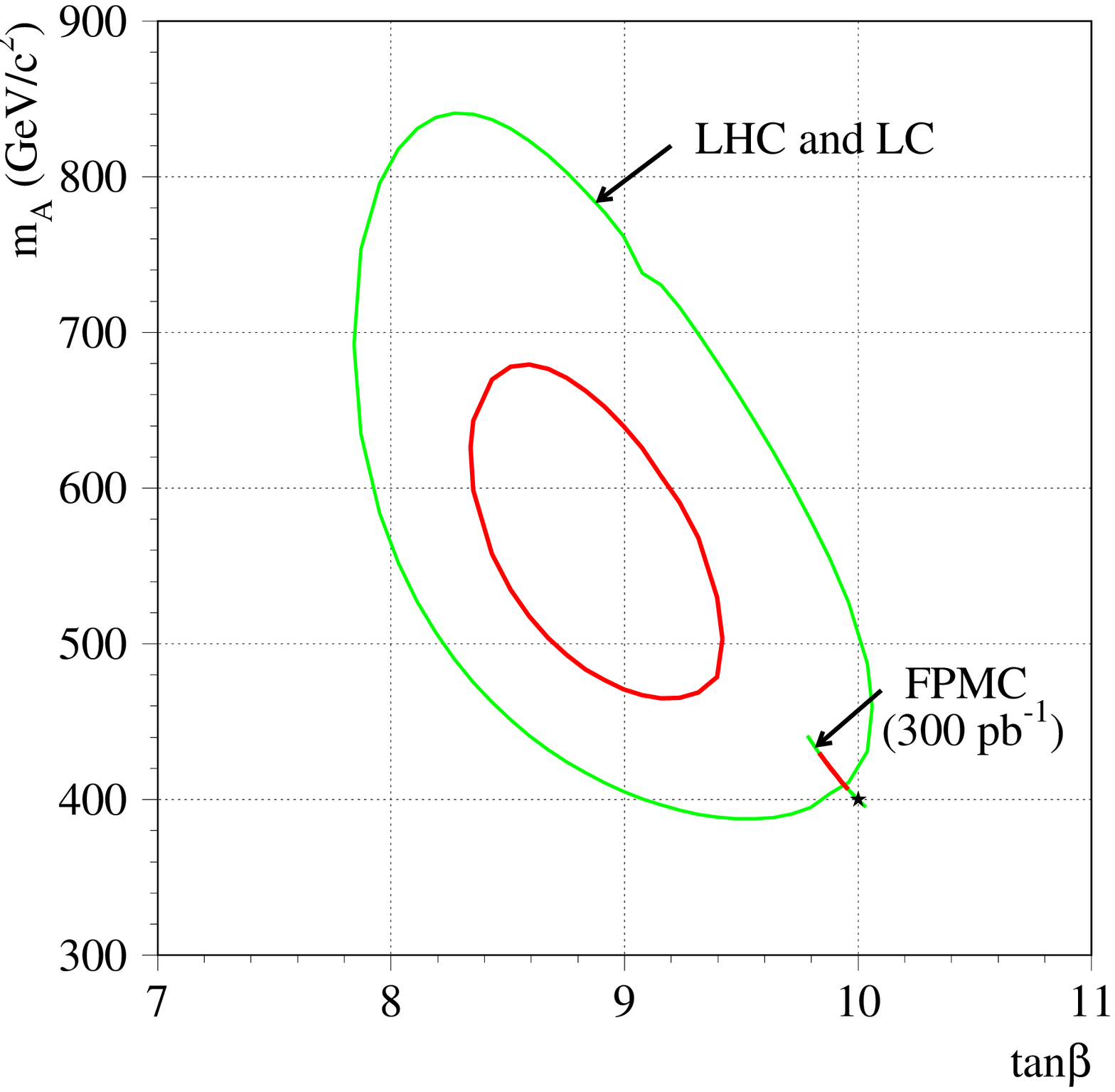}
\includegraphics[width=2.65in]{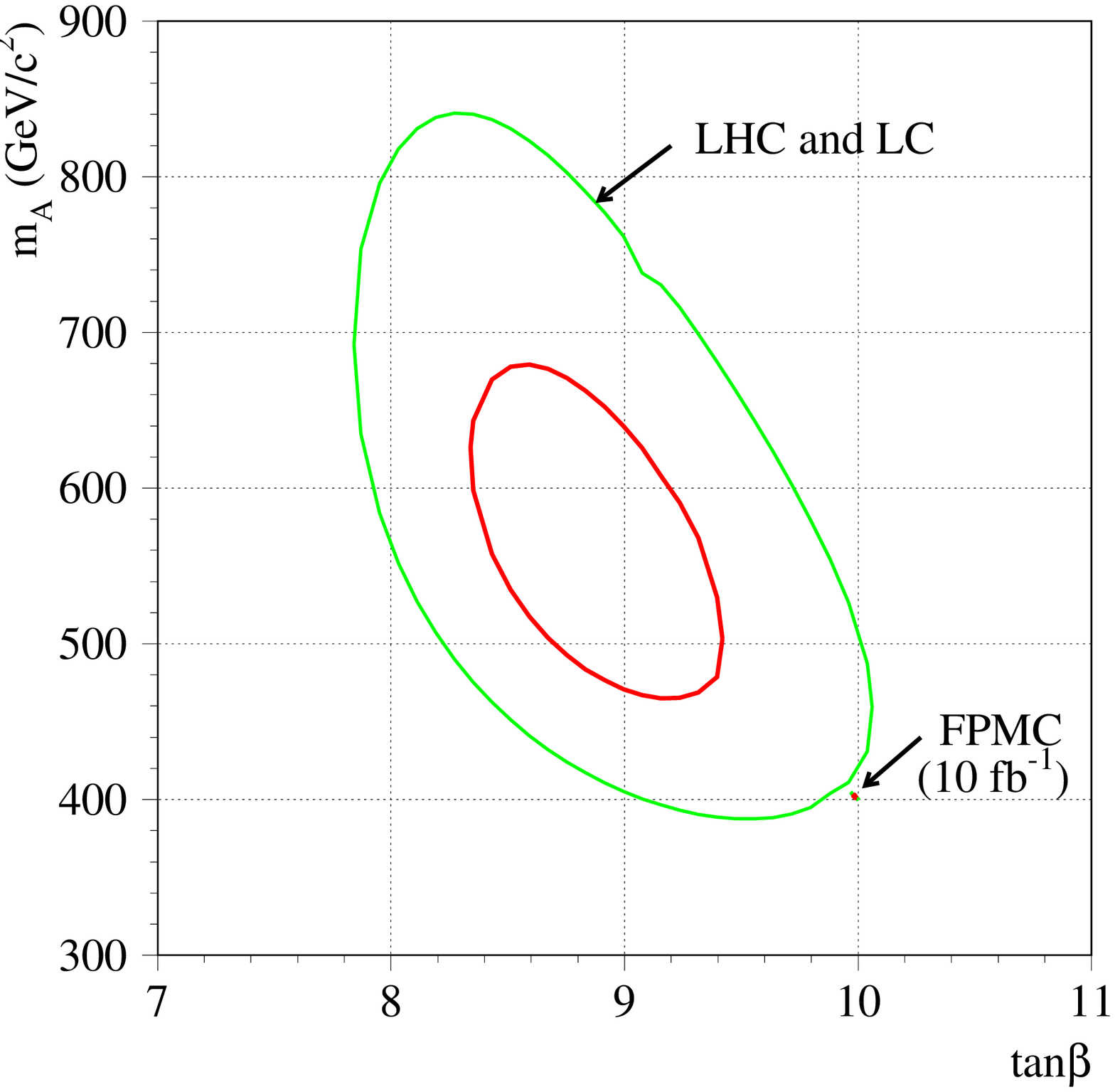}
\vspace*{0.02in}
\includegraphics[width=2.65in]{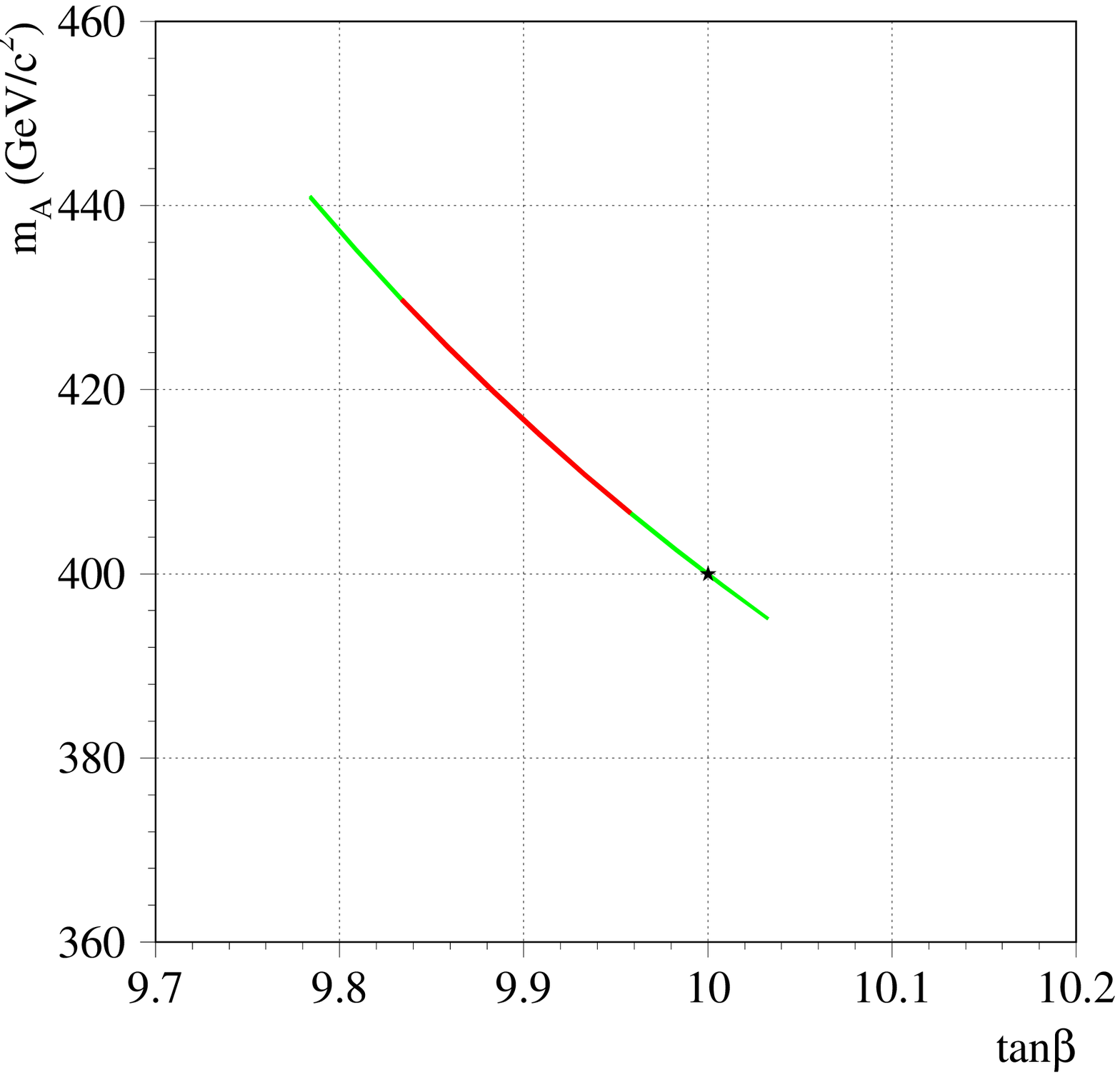}
\includegraphics[width=2.65in]{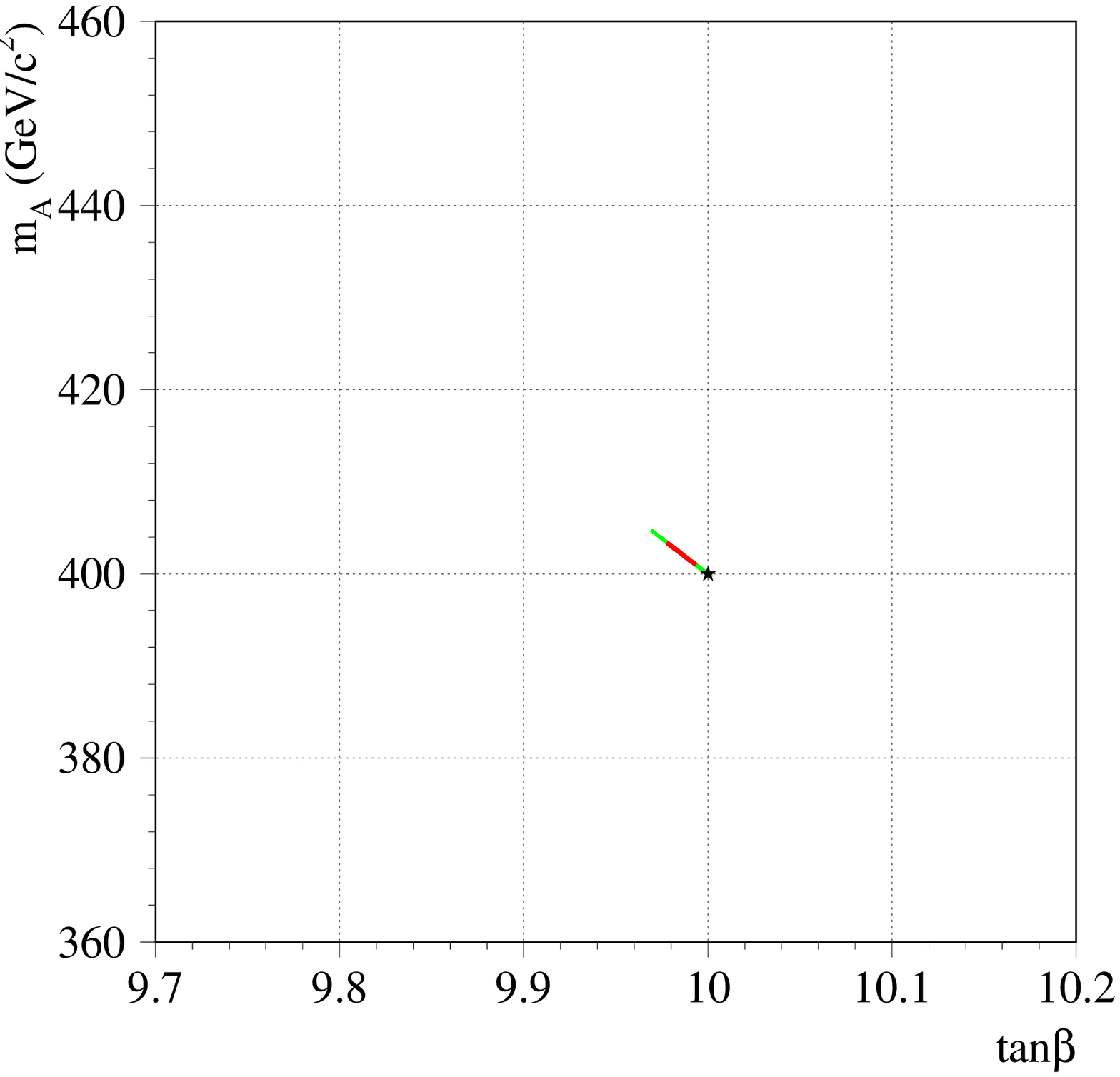}
\caption[The implications of the $\hl$ scan for the MSSM parameter space]
{The implications of the $\hl$ scan for the MSSM $[\mha,\tanb]$
parameter space, assuming all other SUSY parameters are known.
In the lower figures, we illustrate the results that would
emerge were there no systematic theoretical uncertainties
in the $\mhl$ computation in terms of input SUSY parameters.
The experimental error of $\mhl$ at a muon collider would not
significantly broaden this line. The LH (RH) lower figure
shows the extent to which the location along this line would
be fixed by $L=0.1\fbi$ ($L=10\fbi$) muon collider measurements of
$N(\mupmum\to\h\to b\anti b)$ and $\gamhtot$, with the former being
the dominant ingredient given its much smaller error.
In the upper two figures, the restrictions (1 and 2 $\sigma$ ellipses)
 that would emerge from
LHC+LC measurements (including the measurement of $\mhl$
with accuracy of order $\pm 30\mev$) are shown. (Note the
much more coarse scale of the upper figures.) 
These figures are from Ref.~\cite{Autin:1999ci}.
Unfortunately, the systematic error ($\gsim \pm 100\mev$, at best)
expected for the $\mhl$ computation
in terms of the input SUSY parameters will cause the potential 
muon collider lines of the lower figures to turn into ellipses 
similar in size to the LHC+LC ellipses and will increase the
size of the  LHC+LC ellipses significantly.
\label{parameters}}
\end{center}
\end{figure*}

\begin{figure}[h]
\begin{center}
\includegraphics[width=3.5in]{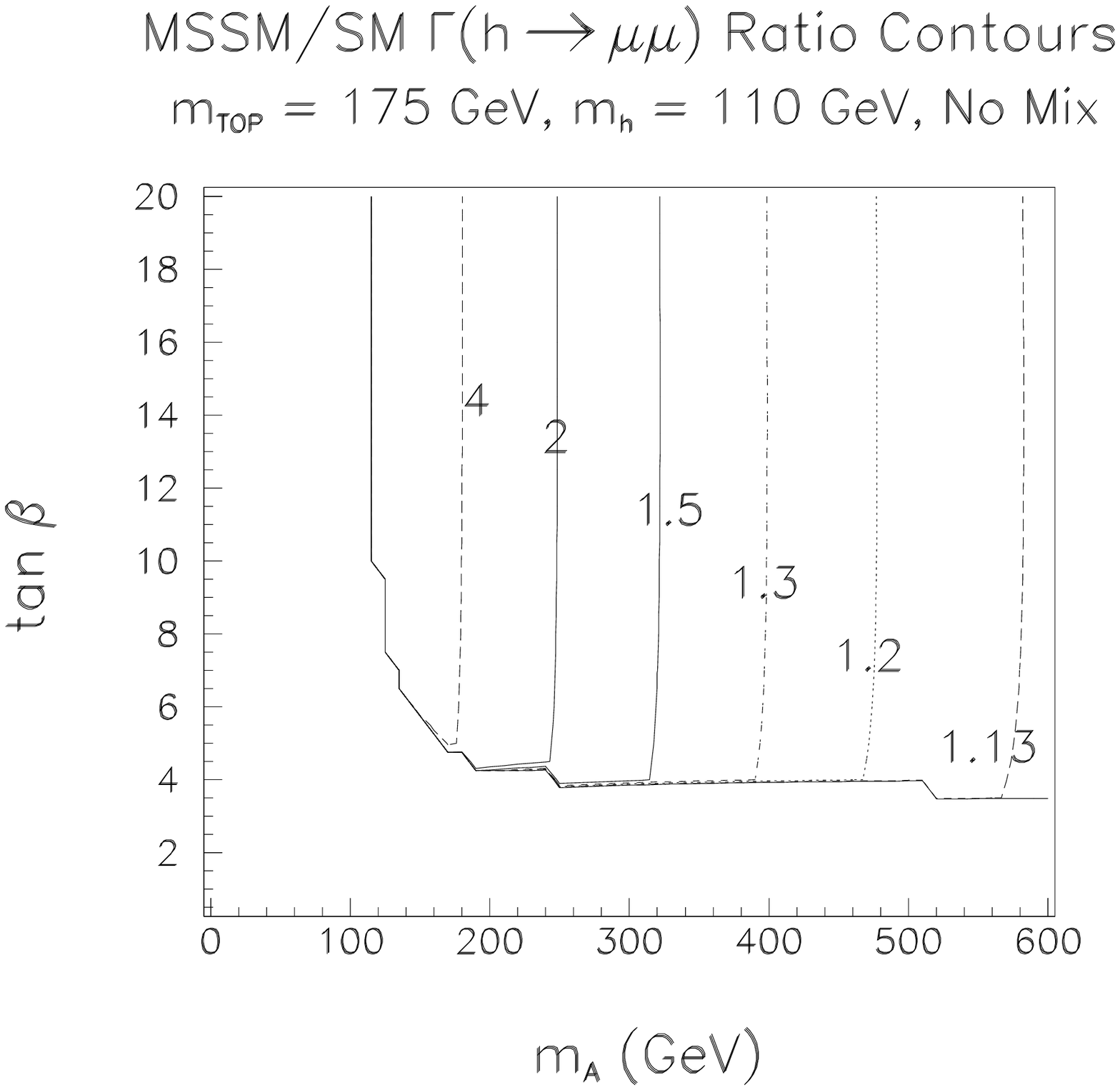}
\caption[]{ Contours in $(\mha,\tanb)$ parameter space
for $\Gamma(\hl\to\mupmum)/\Gamma(\hsm\to\mupmum)$.
We have assumed a no-mixing SUSY scenario and employed $\mhl=\mhsm=110\gev$. 
For maximal mixing, there is little change in the contours --- 
only the size of the allowed range is altered. From \cite{gunionmumu}.}
\label{mumucontours}
\end{center}
\end{figure}

Given the above sensitivity, the next question is the extent to which
parameters of the superymmetric Higgs sector can be determined with
some reasonable level of precision. To study this, an input MSSM
model was assumed with $\mhl=110\gev$, $\mha=400\gev$,
$\tanb=10$ and $A_t=\mu=M_{\rm SUSY}=1\tev$. Various observables
were computed as a function of $\mha$ and $\tanb$. 
Let us for the moment imagine that $\mhl$ can be computed theoretically
with arbitrary accuracy in terms of the input SUSY model
parameters.  Were this the case, then the value of $\mhl$ would
determine $\mha$ as a function of $\tanb$ (or vice versa)
given the fixed SUSY breaking scenario parameters. The $b\anti b$
event rate and, to a lesser extent, $\gamhltot$ determine
the location along the line allowed by
the fixed value of $\mhl$. This line in $[\mha,\tanb]$ parameter
space is illustrated in the lower figures of 
Fig.~\ref{parameters} for the above sample model \cite{Murray:2001es}.
Also shown in these lower figures is the extent to which
experimental measurements of 
$N(\mupmum\to \hl\to b\anti b)$ and $\gamhltot$ 
for $L=0.1\fbi$ and $L=10\fbi$ would restrict
the location along this line.
The accuracy ($\pm [0.1-0.3]\mev$)
with which $\mhl$ can be determined experimentally
at the muon collider would not significantly broaden this line.
For the experimental accuracies of $\pm 90\mev$ at the LHC and 
$\pm 30\mev$ at the LC, the line turns into the ellipses of the upper figures
of Fig.~\ref{parameters}. 
Unfortunately,  due to the expected level of theoretical uncertainties
in the computation of $\mhl$ the muon collider results are certainly
unrealistic and even the LHC+LC ellipses are probably overly optimistic.
We estimate that one might eventually be able to achieve a theoretical
accuracy of $\pm 100\mev$ for the $\mhl$ computation in terms
of the model parameters. (Currently, the accuracy of the theoretical
computations is $\sim \pm [2-3]\gev$, so that much higher-loop work will
be required to reach this level.)  This would be comparable to
the LHC experimental errors on $\mhl$. Thus, the reality may
be that LHC+LC ellipses of the upper half of Fig.~\ref{parameters}
will be substantially enlarged. In any case, the ellipse sizes in both cases
would most probably be determined by the accuracy of the theoretical
computation of $\mhl$ as a function of SUSY parameters.
A determination of the allowed elliptical regions including a reasonable
level of systematic uncertainty for the $\mhl$ computation should be made.
Despite this systematic uncertainty
from the $\mhl$ computation, it is nonetheless
clear that strong constraints would be 
imposed on the allowed regions in the multi-dimensional MSSM parameter space
(that includes $m_A$ and $\tan \beta$ and the SUSY-breaking parameters)
in order to achieve consistency with 
the measurements of $\mhl$, $\overline\sigma(\mupmum\to\hl\to b\anti b)$ 
and $\gamhltot$.

One very important probe of the physics
of a light $\h$ that is only possible at a muon collider is
the possibility of measuring $\Gamma(\h\to\mupmum)$.
Typically, the muon collider data must be combined with LC
and/or LHC data to extract this very fundamental coupling.
If the $\h$ is SM-like then the following determinations are
possible.
\begin{description}
\item{ 1)}\hskip 1in $\Gamma(\h\to\mupmum)={[\Gamma(\h\to\mupmum)\br(\h\to
b\anti b)]_{\mbox{$\mu$C}}\over \br(\h\to b\anti b)_{\rm NLC}}$;
\item{ 2)}\hskip 1in $\Gamma(\h\to\mupmum)={[\Gamma(\h\to\mupmum)\br(\h\to
W\wstar)]_{\mbox{$\mu$C}}\over\br(\h\to W\wstar)_{\rm NLC}}$;
\item{ 3)}\hskip 1in $\Gamma(\h\to\mupmum)={[\Gamma(\h\to\mupmum)\br(\h\to
Z\zstar)]_{\mbox{$\mu$C}}\gamhtot\over\Gamma(\h\to Z\zstar)_{\rm NLC}}$;
\item{ 4)} \hskip 1in $\Gamma(\h\to\mupmum)={[\Gamma(\h\to\mupmum)\br(\h\to
W\wstar)\gamhtot]_{\mbox{$\mu$C}}\over\Gamma(\h\to W\wstar)_{\rm NLC}}$.
\end{description}
Using the above, a determination of $\Gamma(\h\to\mupmum)$ with accuracy
$\pm 4\%$ would be possible for an $L\sim 0.2\fbi$ muon collider run 
on the $\h$ peak and combining with
LC($200\fbi$) data.  In the MSSM context, such precision
means that one would have $3\sigma$ or greater difference
between the expectation for the $\hsm$ vs. the result for the $\hl$
if $\mha\leq 600\gev$, assuming $\mhl\lsim 135\gev$ (the MSSM upper limit).
Further, this is an absolutely direct and model independent determination
of $\Gamma(\hl\to\mupmum)$ that for certain has no systematic
theoretical uncertainties.  Of course, the caveat remains
that there are peculiar MSSM parameter choices for which
`decoupling' occurs very rapidly and the $\hl\to\mupmum$
coupling would be independent of $\mha$. However, we would know
ahead of time from the SUSY spectrum observed at the LHC whether
or not such a peculiar scenario was relevant. 
Finally, we emphasize that the muon collider provides the only accurate
probe of this 2nd generation lepton 
coupling~\footnote{For $\mh\sim 120\gev$, estimates
given M. Battaglia at the recent ECFA/DESY meeting for the accuracy
to which $\Gamma(\hl\to \mupmum)$ can be determined at an $\epem$
collider range from
$\pm 30\%$ for $\rts=350\gev$ running ($L=500\fbi$) (not much different
for $\rts=800\gev$ and $L=1\abi$) to
$\pm 7\%$ for $\rts=3\tev$ ($L=5\abi$).}  and would thus be
one of the best checks of the the SM or MSSM 
explanation of lepton masses.

To summarize,
if a Higgs is discovered at the LHC, or possibly earlier at the Fermilab 
Tevatron, attention will turn to determining  whether this Higgs has the 
properties expected of the Standard Model Higgs. If the Higgs is discovered
at the LHC, it is quite possible that supersymmetric states will be 
discovered concurrently. The next goal for a linear collider or a muon collider
will be to better measure the Higgs boson properties to determine if 
everything is consistent within a supersymmetric framework or consistent
with the Standard Model.
A Higgs factory of even modest luminosity can provide
uniquely powerful constraints on the parameter space of the supersymmetric
model via the highly accurate determination
of the total rate for $\mupmum\to\hl\to b\anti b$ (which has almost
zero theoretical systematic uncertainty due to its insensitivity
to the unknown $m_b$ value), the moderately accurate determination
of the $\hl$'s total width and the remarkably accurate, unique and 
model-independent determination of the $\hl\mupmum$ coupling constant. 

\section{$h\to \tau^+\tau^-$}

A particularly important channel is the $\tautau$ final 
state \cite{Taus}
\begin{equation}
\mumu \to \tautau.
\end{equation}
In the SM at tree level, this $s$-channel process 
proceeds in two ways, via $\gamma/Z$ exchange and Higgs boson 
exchange. The former involves the SM gauge couplings and 
presents a characteristic $FB$ (forward-backward in the scattering
angle) asymmetry and a $LR$ (left-right in beam polarization) asymmetry;
the latter is governed by the Higgs boson couplings to
$\mumu,\tautau$ proportional to the fermion masses
and is isotropic in phase space due to spin-0 exchange.
The unambiguous establishment of the $\tautau$ 
signal would allow a determination of the relative coupling strength 
of the Higgs boson to $b$ and $\tau$ and thus test the usual
assumption of $\tau-b$ unification. The angular distribution
would probe the spin property of the Higgs resonance.

The differential cross section for $\mu^- \mu^+ \to \tau^- \tau^+$ 
via $s$-channel Higgs exchange can be expressed as
\begin {equation}
{{d\sigma_h(\mumu \to h\to\tautau)}\over{d\cos\theta}}
={1\over 2}{\overline \sigma_h}\ (1+P_-P_+)
\label{higgs}
\end{equation}
where $\theta$ is the scattering angle between $\mu^-$ and $\tau^-$,
$P_\mp$ the percentage longitudinal polarizations of the initial 
$\mu^\mp$ beams, with $P=-1$ purely left-handed, $P=+1$ purely 
right-handed and $P=0$ unpolarized.

The differential cross section for the SM background is given by
the $\gamma/Z$ contributions
\begin {equation}
{d\sigma_{SM}^{} \over
d\cos\theta}={3\over8}\sigma_{QED}^{} A
[1-P_+P_-+(P_+-P_-)A_{LR}](1+\cos^2\theta+
{8\over 3} \cos\theta A^{eff}_{FB}).
\label{eff-bkgrnd}
\end{equation}
Here the effective $FB$ asymmetry factor is
\begin {equation}
A^{eff}_{FB}={{A_{FB}+P_{eff}A_{LR}^{FB}} \over {1+P_{eff}A_{LR}}},
\label{eff-FB}
\end{equation}
with the effective polarization 
\begin {equation}
P_{eff}= {{P_+-P_-}\over{1-P_+P_-}},
\label{effP}
\end{equation}
and
\begin{equation}
A_{LR}^{FB}= {{\sigma_{LR + RL \to LR}-\sigma_{LR + RL
\to RL}}\over{\sigma_{LR + RL \to LR}+\sigma_{LR + RL \to RL}}}.
\label{factors}
\end {equation}
$A_{FB},A_{LR}$ are the standard asymmetries.
For the case of interest where initial and final state particles 
are leptons, $A_{LR}=A_{LR}^{FB}$.

\begin{figure}[tb]
\begin{center}
\includegraphics[width=3.5in]{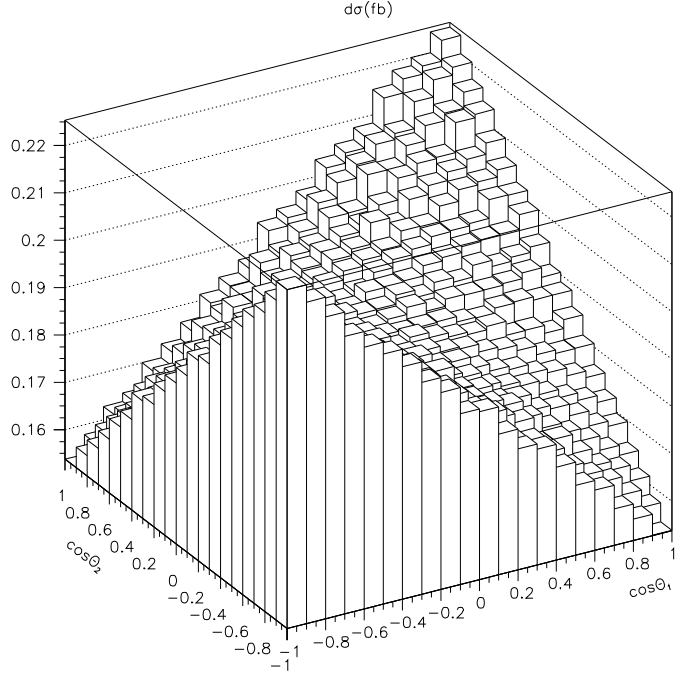}
\caption[]{Double differential distribution for
$\mumu\to h \to \tautau \to \rho^-\nu_\tau\rho^+\bar \nu_\tau$.
$\sqrt s=m_h=120$ GeV is assumed. Initial $\mu^\mp$ beam
polarizations are taken to be $P_-=P_+=0.25$. The Higgs 
production cross section is convoluted
with Gaussian energy distribution for a resolution $R=0.05\%$. 
\label{two}}
\end{center}\end{figure}

From the cross section formulas of 
Eqs.~(\ref{higgs}) and (\ref{eff-bkgrnd}), 
the enhancement factor of the signal-to-background ratio ($S/B$)
due to the beam polarization effects is 
\begin{equation}
{S\over B} \sim {1+P_-P_+\over 1-P_-P_+ +(P_+-P_-)A_{LR}}.
\label{pmu}
\end{equation}

The final state polarization configurations of $\tautau$ from the
Higgs signal and the SM background are very different. 
There is always a charged track to define a 
kinematical distribution for the decay. 
In the $\tau$-rest frame, the normalized 
differential decay rate can be written as
\begin {equation}
{1\over \Gamma}{d\Gamma_i \over {d\cos\theta}}=
{B_i\over 2} (a_i+b_iP_\tau \cos\theta)
\label{tau-decay}
\end {equation}
where $\theta$ is the angle between the momentum direction 
of the charged decay product in the $\tau$-rest 
frame \cite{TauDecay} and the $\tau$-momentum direction, 
$B_i$ is the branching fraction for a given
channel $i$, and $P_\tau=\pm 1$ is the $\tau$ helicity. 
For the two-body decay modes, $a_i$ and $b_i$ are constant and
given by 
\begin{eqnarray}
&& a_\pi=b_\pi=1,\\  
&& a_i=1\quad {\rm and}\quad b_i=-{m_\tau^2-2m_i^2\over m_\tau^2+2m_i^2}\quad
{\rm for}\quad i=\rho,\ a_1^{}.
\end{eqnarray} 
For the three-body leptonic decays, the $a_{e,\mu}^{}$ 
and $b_{e,\mu}^{}$ are not constant for a given three-body 
kinematical configuration
and are obtained by the integration over the energy fraction 
carried by the invisible neutrinos. 
One can quantify the event distribution shape by defining 
a ``sensitivity'' ratio parameter
\begin{equation}
r_i= {b_i\over a_i}.
\label{sens}
\end{equation}
For the two-body decay modes, the sensitivities are
$r_\pi=1,\ r_\rho=0.45$ and $r_{a_1}=0.007$. The $\tau\to a_1\nu_\tau$
mode is consequently less useful in connection with the $\tau$ polarization
study. As to the three-body leptonic modes, although experimentally readily
identifiable, the energy smearing from the decay makes it hard to reconstruct 
the $\tautau$ final state spin correlation. 

\begin{figure}[tb]
\begin{center}
\includegraphics[width=3.5in]{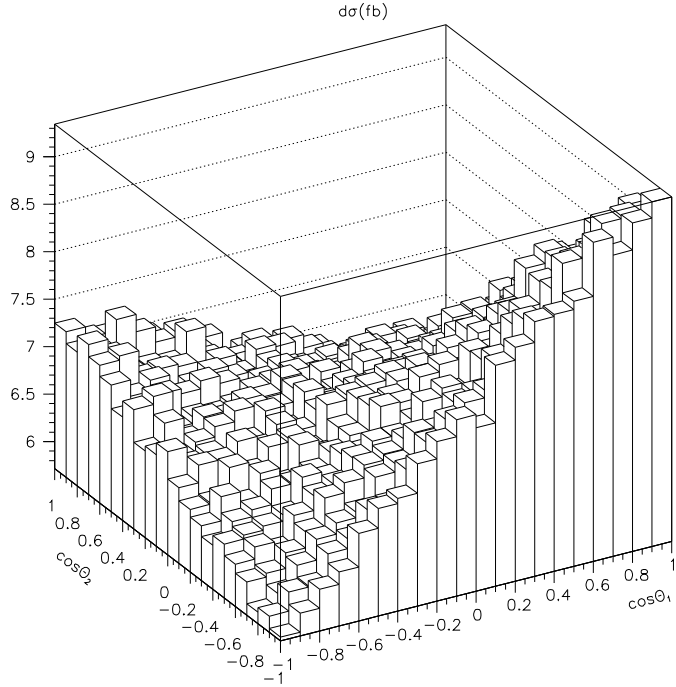}
\vspace{10pt}
\caption[]{Double differential distribution for
$\mumu\to \gamma^*/Z^* \to \tautau \to \rho^-\nu_\tau\rho^+\bar \nu_\tau$.
$\sqrt s=120$ GeV is assumed. Initial $\mu^\mp$ beam
polarizations are taken to be $P_-=P_+=0.25$. The SM
production cross section is convoluted with Gaussian energy 
distribution for a resolution $R=0.05\%$. 
\label{three}}
\end{center}\end{figure}

The differential distribution for the two charged particles
($i,j$) in the final state from $\tautau$ decays 
respectively can be expressed as
\begin{eqnarray}
{d\sigma \over {d\cos\theta_id\cos\theta_j}} 
\sim \sum_{P_\tau=\pm 1} {B_{i}B_{j}\over 4}\ (a_i+b_iP_{\tau^-} \cos\theta_i)
(a_j+b_jP_{\tau^+}\cos\theta_j),
\end{eqnarray}
where $\cos\theta_i\ (\cos\theta_j)$ is defined in 
$\tau^-\ (\tau^+)$ rest frame as in Eq.~(\ref{tau-decay}).
For the Higgs signal channel,  $\tautau$ helicities are correlated
as  $LL\ (P_{\tau^-}=P_{\tau^+}=-1)$ 
and $RR\ (P_{\tau^-}=P_{\tau^+}=+1)$. 
This yields the spin-correlated differential cross section
\begin{eqnarray}
 {d\sigma_h \over {d\cos\theta_id\cos\theta_j}} 
 =  (1+P_-P_+)\sigma_h\ {B_{i}B_{j}\over 4}\ 
[a_ia_j+b_ib_j\cos\theta_i\cos\theta_j],
\label{LL&RR}
\end {eqnarray}
We expect that the distribution reaches
maximum near 
$\cos\theta_i=\cos\theta_j=\pm 1$ and minimum
near $\cos\theta_i=-\cos\theta_j=\pm 1$. 
How significant the peaks are depends on the sensitivity 
parameter in Eq.~(\ref{sens}). Here we simulate the double 
differential distribution of Eq.~(\ref{LL&RR})
for $\mumu\to h \to \tautau \to \rho^-\nu_\tau\rho^+\bar \nu_\tau$
and the result is shown in Fig.~\ref{two}.
Here we take $\sqrt s=m_h=120$ GeV for illustration. 
The Higgs production cross section is convoluted
with Gaussian energy distribution  
for a resolution $R=0.05\%$. 
We see distinctive peaks in the distribution near 
$\cos\theta_{\rho^-}=\cos\theta_{\rho^+}=\pm 1$, 
as anticipated. In this demonstration, we have taken 
$\mu^\mp$ beam polarizations to be $P_-=P_+=25\%$,
which is considered to be natural with little cost to 
beam luminosity.

In contrast, the SM background via $\gamma^*/Z^*$ produces
$\tautau$ with helicity correlation 
of  $LR\ (P_{\tau^-}=-P_{\tau^+}=-1)$ 
and $RL\ (P_{\tau^-}=-P_{\tau^+}=+1)$. 
Furthermore, the numbers of the left-handed and right-handed 
$\tau^-$ at a given scattering angle are different because 
of the left-right asymmetry, so the initial muon beam
polarization affects the $\tautau$ spin correlation non-trivially.
Summing over the two polarization combinations in $\tautau$
decay to particles $i$ and $j$, we have
\begin {eqnarray}
{d\sigma_{SM}^{} \over {d\cos\theta_id\cos\theta_j}} = 
&& (1-P_-P_+)\sigma_{SM}^{}\ (1+P_{eff}A_{LR})\times
\nonumber\\ 
&&{B_{i}B_{j}\over 4}[(a_ia_j-b_ib_j\cos\theta_i\cos\theta_j) + 
A^{eff}_{LR}(a_ib_j\cos\theta_j-a_jb_i\cos\theta_i)].
\label{LR&RL}
\end {eqnarray}
The final state spin correlation for $\mumu\to \gamma^*/Z^*\to \tautau$ 
decaying into $\rho^-\rho^+$ pairs is shown in Fig.~\ref{three}.
The maximum regions near
$\cos\theta_{\rho^-}=-\cos\theta_{\rho^+}=\pm 1$ 
are clearly visible. Most importantly, the peak regions in
Figs.~\ref{two} and \ref{three} occur exactly in the opposite positions 
from the Higgs signal. We also note that the spin correlation from 
the Higgs signal is symmetric, while that from the background is
not. The reason is that the effective $LR$-asymmetry in the background
channel changes the relative weight of the two maxima, which becomes
transparent from the last term in Eq.~(\ref{LR&RL}). 

\begin{figure}[tb]
\begin{center}
\includegraphics[width=3.5in]{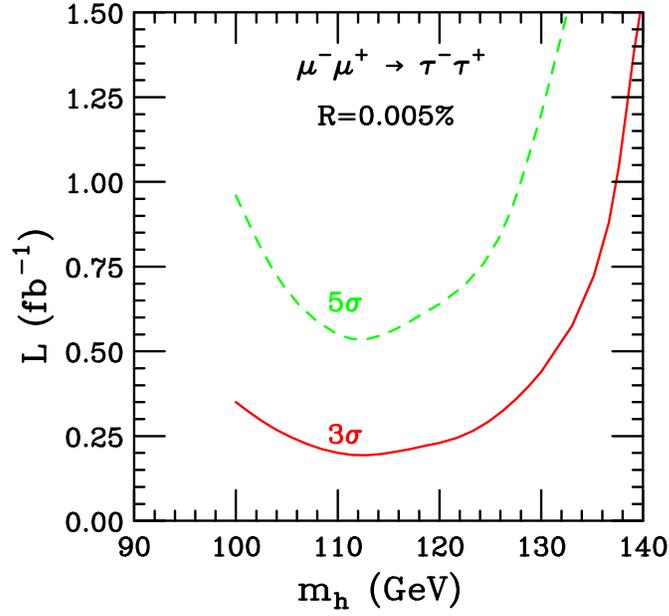}
\vspace{10pt}
\caption[]{Integrated luminosity (in $\fbi$) needed for 
observing the two-body decay channels $\tau \to \rho\nu_\tau$ 
and $\tau \to \pi\nu_\tau$ at $3\sigma$ (solid) and $5\sigma$ 
(dashed) significance. 
Beam energy resolution $R=0.005\%$ and a $25\%$ 
polarization are assumed.
\label{lum}}
\end{center}\end{figure}

We next estimate the luminosity needed for signal observation
of a given statistical significance. The results are shown in 
Fig.~\ref{lum}. The integrated luminosity ($L$ in $\fbi$) 
needed for observing the characteristic two-body decay channels 
$\tau \to \rho\nu_\tau$ and $\tau \to \pi\nu_\tau$ 
at $3\sigma$ (solid) and $5\sigma$ (dashed) significance
is calculated for both signal and SM background with
$\sqrt s=m_h$. Beam energy resolution $R=0.005\%$ and a $25\%$
$\mu^\pm$ beam polarization are assumed. 

We estimate
the statistical error on the cross section measurement. If we
take the statistical error to be given by
\begin{equation}
\epsilon = {\sqrt{S+B}\over S} = 
{1\over \sqrt L}\ {\sqrt{\sigma_S+\sigma_B} \over \sigma_S},
\end{equation}
summing over both $\rho\nu_\tau$ and $\pi\nu_\tau$ channels 
for $R=0.005\%$, a $25\%$ beam polarization with 1 $\fbi$ 
luminosity, we obtain
\begin{equation}
\begin{array}{lcccc}
\sqrt s = m_h\ (\gev)\quad& 100  & 110 & 120 & 130 \\
\epsilon\  (\%)\quad      &  27  &  21 &  23 &  32 
\end{array}
\label{eps}
\end{equation}
The uncertainties on the cross section measurements determine 
the extent to which the $h\tautau$ coupling can be measured.

In summary, we have demonstrated the feasibility of observing
the resonant channel $h\to \tautau$ at a muon collider.
For a narrow resonance like the SM Higgs boson, 
a good beam energy resolution is crucial for a clear signal. 
On the other hand, a moderate beam polarization would not help
much for the signal identification. The integrated luminosity
needed for a signal observation is presented in Fig.~\ref{lum}.
Estimated statistical errors for the $\mumu \to h \to \tautau$
cross section measurement are given in Eq.~(\ref{eps}).
We emphasized the importance of final state spin correlation
to purify the signal of a scalar resonance and to confirm
the nature of its spin. It is also important to carefully
study the $\tautau$ channel of a supersymmetric Higgs
boson which would allow a determination of the relative
coupling strength of the Higgs to $b$ and $\tau$.

\section{Heavy Higgs Bosons}

As discussed in the previous section,
precision measurements of the light Higgs boson properties might make it 
possible to detect deviations with respect to expectations for a SM-like
Higgs boson that would point to 
a limited range of allowed masses for the heavier Higgs bosons.
This becomes more difficult in the decoupling limit where the differences 
between a supersymmetric and Standard Model Higgs are smaller. Nevertheless 
with sufficiently precise measurements of the Higgs branching fractions, it 
is possible that the heavy Higgs boson masses can be inferred. 
A muon collider light-Higgs factory might be essential in this process.

In the context of the MSSM, 
$\mha$ can probably~\footnote{For the peculiar
parameter regions with `early' decoupling, mentioned earlier,
this would not be possible.  However, as noted earlier,
the SUSY spectrum observed at the LHC would allow us
to determine if we are in such an exceptional region of parameter
space.} be restricted to
within $50\gev$ or better if $\mha<500\gev$.
This includes the $250-500\gev$
range of heavy Higgs boson masses for which discovery is not possible 
via $\hh\ha$ pair production 
at a $\rts=500\gev$ LC. Further, the $\ha$ and $\hh$
cannot be detected in this mass range at either the LHC or LC 
for a wedge of moderate $\tanb$ values. (For large enough 
values of $\tanb$ the heavy Higgs bosons are expected to be observable
in $b\anti b \ha,b\anti b \hh$ production
at the LHC via their $\tau ^+\tau ^-$ decays and also at the LC.)

A muon collider can fill some, perhaps all of this moderate $\tanb$ wedge.
If $\tanb$ is large the $\mupmum \hh$ and $\mupmum\ha$ couplings (proportional
to $\tanb$ times a SM-like value) are enhanced,
thereby leading to enhanced production rates in $\mupmum$ collisions.
The most efficient procedure is the operate the muon collider
at maximum energy and produce the $\hh$ and $\ha$ (often as overlapping
resonances) 
via the radiative return mechanism. By looking for a peak
in the $b\anti b$ final state, the $\hh$ and $\ha$ 
can be discovered and, once discovered, the machine $\rts$
can be set to $\mha$ or $\mhh$ and factory-like precision studies pursued.
Note that the $\ha$ and $\hh$ are typically broad enough that $R=0.1\%$
would be adequate to maximize their $s$-channel production rates.
In particular, $\Gamma\sim 30$~MeV
if the $t\overline{t}$ decay channel is not open, and $\Gamma\sim 3$~GeV if it
is. Since $R=0.1\%$ is sufficient, much higher luminosity
($L\sim 2-10~{\rm fb}^{-1}
/{\rm yr}$) would be possible as compared to that 
for $R=0.01\%-0.003\%$ as required for studying the $\hl$.
 
In short, for those portions of parameter space characterized
by moderate $\tanb$ and $\mha\gsim 250\gev$,
which  are particularly difficult for both
the LHC and the LC, the muon collider would be the only 
place that these extra Higgs bosons can be discovered and their properties 
measured very precisely.~\footnote{The $\gam\gam$ collider option
at an LC would also allow $\hh,\ha$ discovery throughout much of
the wedge region \cite{ourstudy}, 
but only the muon collider could directly scan
for their total widths and determine their $\mupmum$ coupling.}

\begin{figure*}[hbt!]
\begin{center}
\includegraphics[width=4in]{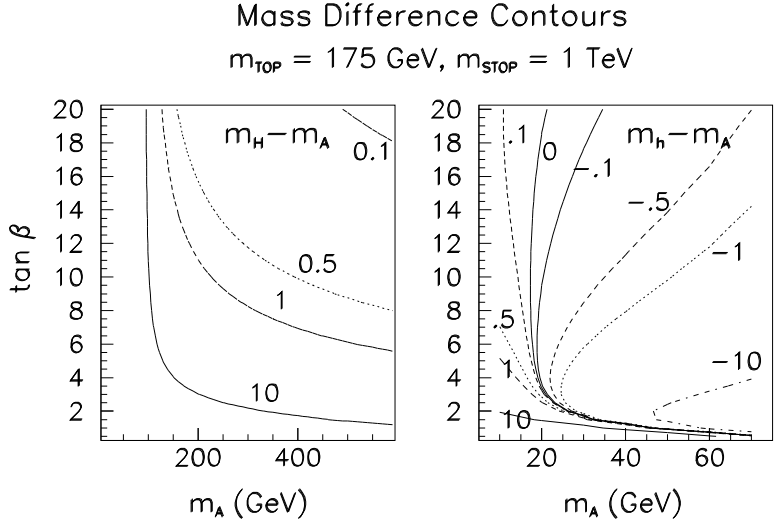}

\caption{Contours of $m_H-m_A$ (in GeV)
in the $(m_H,\tan \beta)$ parameter space. Two-loop/RGE-improved
radiative corrections are included taking $m_t=175$~GeV, $m_{\tilde{t}}=1$~TeV,
and neglecting squark mixing. 
\label{delmhiggs}}
\end{center}\end{figure*}

\begin{figure*}[hbt!]
\begin{center}
\includegraphics[width=4in]{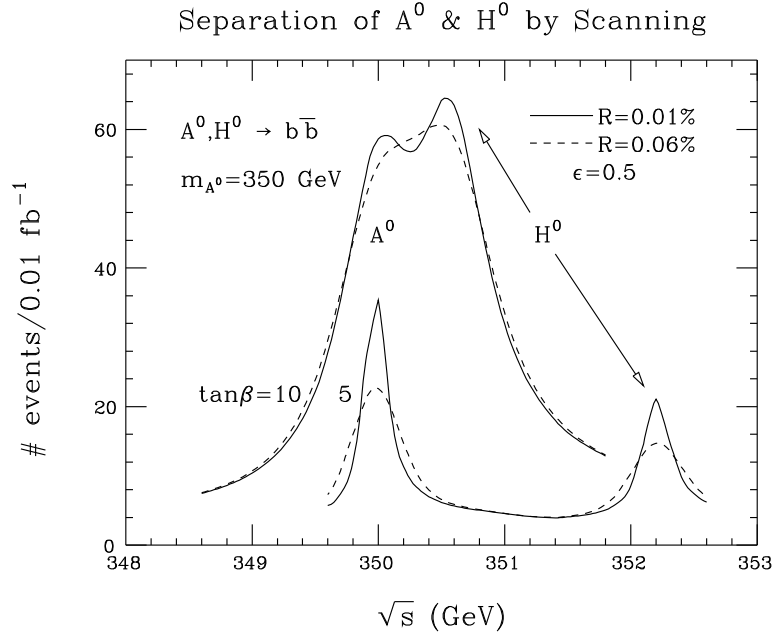}

\caption[Separation of $A$ and $H$ signals for $\tan\beta=5$ and $10$]
{Separation of $A$ and $H$ signals for $\tan\beta=5$ and $10$. From  
Ref.~\cite{Barger:1997jm}. \label{H0-A0-sep}}
\end{center}\end{figure*}

In the MSSM, the heavy Higgs bosons are largely degenerate, especially in the 
decoupling limit where they are heavy. Large values of $\tan \beta$ heighten
this degeneracy as shown in Fig.~\ref{delmhiggs}. 
A muon collider with sufficient energy resolution might be
the only possible means for separating out these states.
Examples showing the $H$ and $A$ resonances for $\tan \beta =5$ and $10$
are shown in Fig.~\ref{H0-A0-sep}. For the larger value of 
$\tan \beta$ the resonances are clearly overlapping. For the better energy 
resolution of $R=0.01\%$, the two distinct resonance peaks are still 
visible, but they are smeared out and merge into one broad
peak for $R=0.06\%$. 

A precise measurement on the heavy Higgs boson masses could provide a powerful
window on radiative corrections in the supersymmetric Higgs 
sector\cite{Berger:2001et}.
Supersymmetry with gauge invariance in the MSSM implies the mass-squared sum
rule
\begin{eqnarray}
&&m_h^2+m_H^2=m_A^2+m_Z^2+\Delta \;,
\end{eqnarray}
where $\Delta $ is a calculable radiative correction (the tree-level sum 
rule results from setting $\Delta =0$). Solving for the mass difference
\begin{eqnarray}
&&m_A-m_H={{m_h^2-m_Z^2-\Delta}\over {m_A+m_H}}\;,
\end{eqnarray} 
one obtains a formula involving observables that can be precisely measured.
For example the error on the $m_Z$ is just 2.2~MeV from the LEP 
measurements\cite{Groom:2000in}, and the light Higgs mass can be measured to
less than an MeV in the $s$-channel. 
The masses of and the mass difference between
the heavy Higgs states $H$ and $A$ can also
be measured precisely by $s$-channel production. The ultimate
precision that can be obtained on the masses of the $H$ and $A$ depends 
strongly on the masses themselves and $\tan \beta$. But a reasonable 
expectation is that a scan through the resonances should be able to 
determine the masses and the mass-difference to some tens of 
MeV\cite{Berger:2001et}.
Altogether these mass 
measurements yield a prediction for the radiative correction $\Delta$ which
is calculable in terms of the self-energy diagrams of the Higgs 
bosons\cite{Berger:1990hg}. To fully exploit this
constraint might, however, prove difficult given the notorious difficulty
of computing Higgs boson masses to high enough loop order
that accuracy better than even a GeV can be achieved.

Finally it will be especially interesting to measure the branching ratios
of these heavy Higgs bosons and compare to the theoretical predictions. 
For $\tan \beta \gtap 5$ the $\hh,\ha$ 
decay more often into $b\overline{b}$ than
into $t\overline{t}$. There is a substantial range of parameter space where
significant numbers of events involving both types of decays will be seen
and new type of determination of $\tanb$ will be possible.
If supersymmetric particle masses are below $\sim \mha/2$, then
the branching ratios for $\ha,\hh$ decays to the many distinguishable
channels provide extremely powerful constraints on the 
soft-supersymmetry-breaking parameters of the 
model \cite{Gunion:1997cc,Gunion:1996qd,Feng:1997xv}.

\section{Higgs Threshold Measurement}
The mass, width and spin 
of a SM-like Higgs boson can also be determined by operating either
a muon collider or a linear collider at the $Z\h$ production threshold.
The rapid rise in the production near the threshold is sensitive to the Higgs 
mass\cite{Barger:1997pv}. Furthermore the spin of the Higgs boson can be 
determined by examining the rise in the cross section near threshold. However,
these measurements require tens of inverse femtobarns to provide a useful 
measurement of the mass $(< 100)$~MeV. These threshold measurements can be 
performed at a LC; with 100~fb$^{-1}$ of integrated luminosity, an error
of less than $100$~MeV can be achieved\cite{Barger:1997pv} for $\mh<150$~GeV. 
This is 
comparable to the other methods at energies above threshold. The only means
to reduce the experimental error on the Higgs mass further to below 1~MeV is to
produce the Higgs in the $s$-channel at a muon collider.

The shape of the $\ell^+\ell^-\to Z\h$ 
threshold cross section can also be used to determine the 
spin and to check the 
CP=+ property of the Higgs\cite{Miller:2001bi}. 
These threshold measurements become of interest for a muon collider in the
case where at least a hundred inverse femtobarns of luminosity is available.

\section{Non-exotic, Non-Supersymmetric SM Higgs Sector Extensions}

Although the standard interpretation of precision electroweak
data is that there should be a light Higgs boson with SM-like
$VV$ couplings, alternative Higgs sector models can be
constructed in which a good fit to the precision data is
obtained even though the Higgs boson with large $VV$ coupling is
quite heavy ($\sim 1\tev$).  The simplest such model \cite{Chankowski:2000an}
is based upon the CP-conserving general two-Higgs-doublet model.
The large $\Delta S>0$ and $\Delta T<0$ coming from the heavy
Higgs with large $VV$ coupling is compensated by an
even larger $\Delta T>0$
coming from a small ($\Delta M\sim 1\gev$ is sufficient) mass splitting
between the $\hpm$ and the other heavy neutral Higgs boson.
The result is a shift in the $\Delta S>0$, $\Delta T>0$ direction
(relative to the usual $\mhsm\sim 100\gev$ scenario in the SM)
that remains well within the current 90\% CL ellipse in the $S,T$ plane.
The first signal for this type of scenario
would be discovery of a heavy SM-like Higgs boson at the LHC.
If such a heavy SM-like Higgs is discovered,
Consistency with precision electroweak data would then require
the above type of scenario or some other exotic new physics scenario.

Models of this type cannot arise in the supersymmetric context
because of constraints on the Higgs self couplings coming
from the SUSY structure.  They require a special `non-decoupling'
form for the potential that could  arise in models
where the two-doublet Higgs sector  is an effective low
energy description up to some scale $\Lambda$ of order 10 TeV or so.
For these special potential forms, 
there is typically also a  Higgs boson $\what h$ 
($\what h=$decoupled-$\hl$ or $\what h=\ha$) with $m_{\what h}<500\gev$
and no tree-level $VV$ coupling. It's primary decay modes would be
to $b\anti b$ or $t\anti t$ (depending upon its mass) and its 
$\mupmum$ coupling would be proportional to $\tanb$.  For a substantial
range of $\tanb$, this $\what h$ could not 
be detected at either the LHC or the LC \cite{Chankowski:2000an}. 
In particular, at the LC even the $\epem\to\zstar\to Z\what h\what h$
process (the quartic coupling being of guaranteed strength)
would only allow $\what h$ discovery up to $150\gev$ ($250\gev$)
for $\rts=500\gev$ ($800\gev$) \cite{farrisgunion}. 

The muon collider could be the key to discovering such a $\what h$.
By running at high energy, the radiative return
tail for $E_{\mupmum}$ might result in production of a detectable
number of events. 
In particular, if $\tanb>5$, operation at maximal $\rts$ with
$R=0.1\%$ would guarantee that  
the $\what h$  would be detected as a $4\sigma$ or higher
bump in the bremsstrahlung tail of 
the $m_{b\anti b}$ distribution after 3 to 4 years
of running. Alternatively, a scan
could be performed to look for the $\what h$.
The scan procedure depends upon how $\Gamma_{\what h}^{\rm tot}$ 
depends on $m_{\what h}$ in that one must always have $R$
such that $\srts\lsim \Gamma_{\what h}^{\rm tot}$; the luminosity
expected for the required $R$ must then be employed.
Further, one must use steps of size $\Gamma_{\what h}^{\rm tot}\sim \srts$.
For $2m_t>m_{\what h}>150\gev$, $\what h\to b\anti b$ and 
$\Gamma_{\what h}^{\rm tot}\sim 0.05-0.1\gev$ 
unless $\tanb<1$.
For $m_{\what h}>2m_t$, $\Gamma_{\what h}^{\rm tot}$ 
rises to at least 1 GeV.
As result, it would be possible to employ
$R=0.05-0.1\%$ or so for $m_{\what h}<2m_t$ rising to
$R=0.5-1\%$ for $m_{\what h}>2m_t$.
In a $3-4$ year program, using earlier quoted nominal yearly $L$'s
for such $R$'s as function of $\rts$, we could imagine devoting:
\bit
\item $L=0.003\fbi$ to 2000 points separated by $0.1\gev$
in $\rts=150-350\gev$ range --- the total luminosity required
would be $L=4\fbi$ or about  3 years of operation.
One would find ($4\sigma$ level) the $\what h$ in the
$b\anti b$ state if $\tanb\gsim 4-5$.
\item $L=0.03\fbi$ to each of 100 points separated by $0.5\gev$ in the
$\rts=350-400\gev$ range --- the corresponding total luminosity used is
$L=3\fbi$ or about 1/2 year of operation.
For $\tanb>6$ ($<6$), one would find the $\what h$ in $b\anti b$ ($t\anti t$)
final state.
\item $L=0.01\fbi$ to each of 100 points separated by $1\gev$ in
the $\rts=400-500\gev$ range --- the total luminosity
employed would be $L=1\fbi$,  or about 1/10 year.
For $\tanb>7$ ($<8$), one would detect the $\what h$ in 
the $b\anti b$ ($t\anti t$) final state.
\eit
In this way, the muon collier would detect the $\what h$  if
$m_{\what h}<2m_t$ and $\tanb\gsim 5$ or if $m_{\what h}>2m_t$ 
for any $\tanb$.
Once discovered, $\rts=m_{\what h}$ could be chosen for the muon
collider and it would be possible to study the $\what h$
properties in detail.

\section{CP Violation}

A muon collider can probe the CP properties of a Higgs boson produced in 
the $s$-channel. One can measure correlations in the $\tau^+\tau^-$ final state
or, if the Higgs boson is sufficiently heavy, in the $t\overline{t}$ 
final state~\cite{Grzadkowski:1995rx,Asakawa:2001es}. In the MSSM at tree-level
the Higgs states $\hl$, $\hh$, and $\ha$ are CP eigenstates, 
but it has been noted
recently that sizable CP violation is possible
in the MSSM Higgs sector through loop 
corrections involving the third generation 
squarks\cite{Pilaftsis:1998dd,Pilaftsis:1998pe}.
As noted earlier, in the MSSM the two heavy neutral Higgs bosons
($\hh$ being CP-even and $\ha$ being CP-odd) are 
almost degenerate with a  mass splitting comparable or less than their widths. 
If there are CP-violating phases in the neutral
Higgs potential, these will cause these CP eigenstates to mix.
The resulting mass splitting between the eigenstates 
can be larger than their widths. 
The excellent mass resolution at the muon collider
would make it possible separate the masses of the $\hh$ and $\ha$ 
bosons.  The measured mass difference could be combined  with 
the mass sum rule to provide a powerful 
probe of this physics. As already noted, various CP asymmetries in the 
$t\overline{t}$ final state can be observed as well, and 
a muon collider is an ideal place to look for these 
effects~\cite{Grzadkowski:1995rx,Asakawa:2001es}.

The most ideal means for determining the CP nature of
a Higgs boson at the muon collider is to employ transversely polarized
muons. For $\h$ production at a muon collider  with muon coupling
given by the form $\anti\mu(a+ib\gamma_5)\mu \h$,
the cross section takes the form
\bea
\overline\sigma_{\h}(\zeta)&=&\overline\sigma_{\h}^0\left(
1+P_L^+P_L^-+P_T^+P_T^-\left[{a^2-b^2\over a^2+b^2}\cos\zeta-{2ab\over a^2+b^2}\sin\zeta \right]
\right)\nonumber\\
&=&\overline\sigma_{\h}^0
\left[1+P_L^+P_L^-+P_T^+P_T^-\cos(2\delta+\zeta)\right]\,,
\eea
where $\overline \sigma_\h^0$ is 
the polarization average convoluted cross section,
$\delta\equiv \tan^{-1}{b\over a}$, 
$P_T$ ($P_L$) is the degree of transverse (longitudinal)
polarization, and $\zeta$ is the 
angle of the $\mu^+$ transverse polarization relative
to that of the $\mu^-$ as measured using the the direction of the $\mu^-$'s
momentum as the $\hat z$ axis. Of course, if there
is no $P_T$ there would be sensitivity to 
$\overline\sigma_{\h}^0\propto a^2+b^2$ only.
Only the $\sin\zeta$ term is truly CP-violating, but the $\cos\zeta$ term
also provides significant sensitivity to $a/b$.
Ideally, one would isolate
${a^2-b^2\over a^2+b^2}$  and ${-2ab\over a^2+b^2}$ by
running at fixed $\zeta=0,\pi/2,\pi,3\pi/2$ and measuring the
asymmetries
(taking $P_T^+=P_T^-\equiv P_T$ and $P_L^\pm=0$)
\bea
{\cal A}_I&\equiv& {\overline\sigma_{\h}(\zeta=0)-\overline\sigma_{\h}(\zeta=\pi) 
\over \overline\sigma_{\h}(\zeta=0)+\overline\sigma_{\h}(\zeta=\pi) }=
P_T^2{a^2-b^2\over a^2+b^2}=P_T^2\cos2\delta\,,\nonumber\\
{\cal A}_{II}&\equiv& {\overline\sigma_{\h}(\zeta=\pi/2)-\overline\sigma_{\h}(\zeta=-\pi/2) 
\over \overline\sigma_{\h}(\zeta=\pi/2)+\overline\sigma_{\h}(\zeta=-\pi/2) }
=-P_T^2{2ab\over a^2+b^2}=-P_T^2\sin2\delta\,.
\nonumber
\eea
If $a^2+b^2$ is already well determined, and the background is known,
then the fractional error in these asymmetries cam be approximated as
${ \delta{\cal A}\over{\cal A}}\propto P_T^2 \sqrt L$,
which points to the need for the highest possible transverse polarization,
even if some sacrifice in $L$ is required. 

Of course, in reality the precession 
of the muon spin in a storage ring makes running at fixed $\zeta$ impossible.
A detailed study is required \cite{Grzadkowski:2000hm}. We attempt
a brief outline. Taking $\vec B=-B\what y$ we may write 
\bea
s_{\mu^-}&=&P_H^-\left[\gamma(\beta,\what z)\cos\theta^- - 
(0,\what x)\sin\theta^-\right]+
P_V^-(0,\what y)\,,\quad
\label{smum}
s_{\mu^+}=P_H^+\left[\gamma(\beta,-\what z)\cos\theta^+ - 
(0,\what x)\sin\theta^+\right]+
P_V^+(0,\what y)\,.
\label{smup}
\eea
Here, $\what z$ is the direction of the $\mu^-$ instantaneous momentum,
$P_H$ ($P_V$) is the horizontal (vertical, i.e. $\what y$) degree of
polarization, $P_H^\pm\cos\theta^\pm=P_L^\pm$, and $\sqrt{[P_H^\pm
\sin\theta^\pm]^2+[P_V^\pm]^2}=P_T^\pm$. For any setup for
initial insertion into the storage ring, $\theta^\pm$
can be computed as functions of the turn number $N_T$
(counting starting with $N_T=1$ the first time the bunch passes the IP).
(For example, if the $\mu^\pm$ beams enter
the storage ring with $\what P_H^\pm=\what p_{\mu^\pm}$, then
$\theta^\pm(N_T) =\omega (N_T-1/2)$, where
$\omega=2\pi \gamma {g_\mu-2\over 2}$, with $\gamma=E/m_\mu$.)
As a function of
$\thm$ and $\thp$, defining $\cm\equiv \cos \thm$ etc.,          
\bea
{\overline\sigma_{\h}(\thp,\thm)\over\overline\sigma_{\h}^0}&=&
(1+\php\phm\cp\cm)+\cos 2\delta(\pvp\pvm+\php\phm\sp\sm)
+{\sin2\delta}(\phm\pvp\sm-\php\pvm\sp)\,.
\label{sigtheta}
\eea
This formula shows that by following the dependence of 
$\overline\sigma_{\h}(\thp,\thm)$ on $N_T$, one can extract 
values for $\cos2\delta$ and $\sin2\delta$. In practice,
it is best to run in several configurations.
To approximate the $\zeta=0$ configuration, one would choose
$\php=\phm=P_H=0.05$, $\thm=\thp$, $\pvp=\pvm=\sqrt{P^2-P_H^2}$. 
To approximate the $\zeta=\pi$ configuration,  choose
$\php=\phm=P_H=0.05$, $\thm=\thp+\pi$, $\pvm=-\pvp=-\sqrt{P^2-P_H^2}$.
To emphasize the $\zeta=\pi/2$ and $\zeta=3\pi/2$ configurations
over many turns of the bunches, we choose
$\phm=P$ ($\pvm=0$), $\php=P_H=0.05$ and $\pvp=\sqrt{P^2-P_H^2}$.
To obtain an accurate measurement of $\delta$, it is necessary to 
develop a strategy for maximizing $\langle P_T^2\rangle \sqrt L$ by selecting
only energetic muons to accelerate and combining bunches. 
Lack of space prevents a detailed description.

\begin{figure}[h!]
\begin{center}
\includegraphics[width=7in]{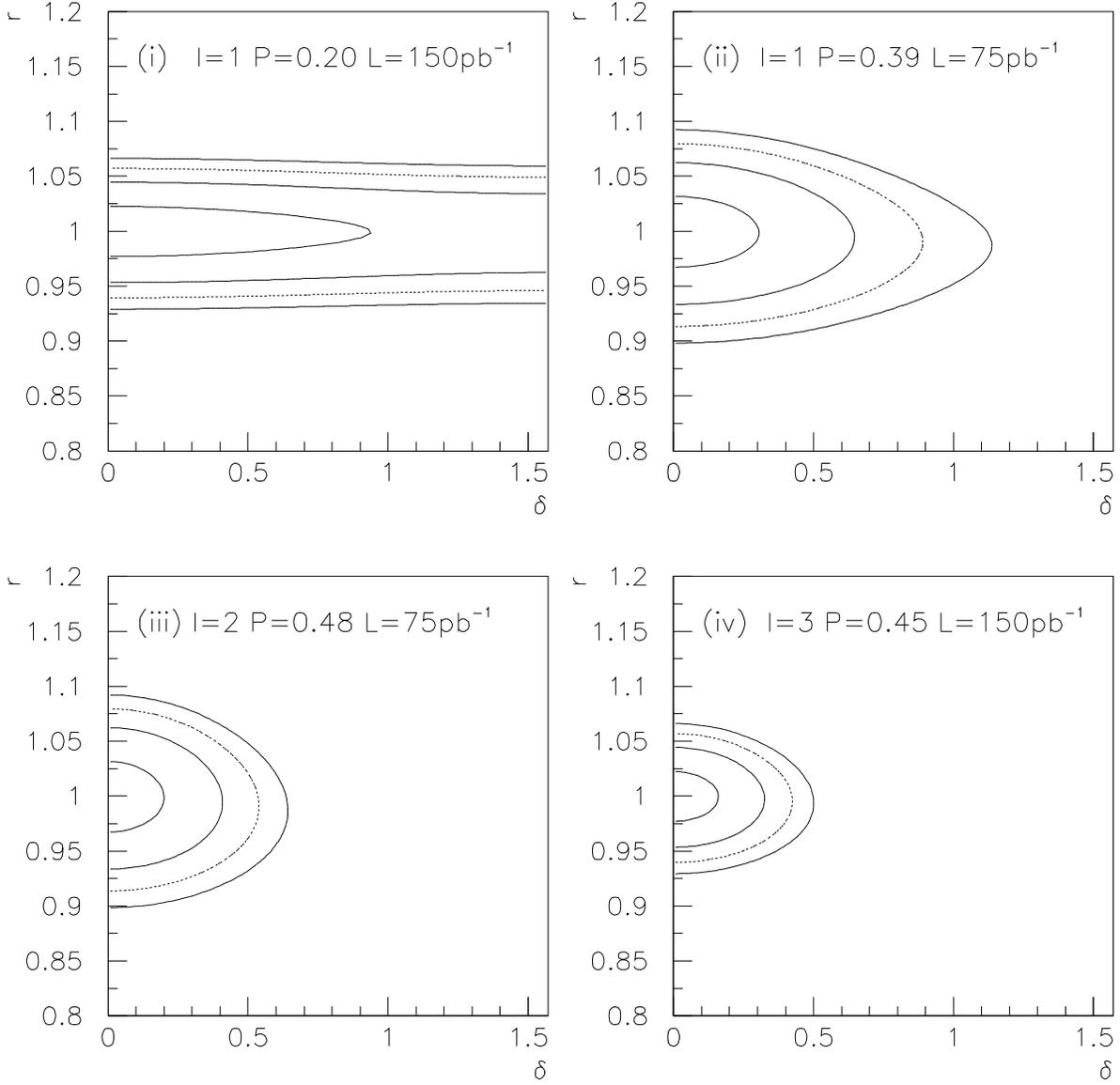}
\caption{Contours at $\Delta\chi^2=1,4,6.635,9$ for the $\hat a$ and $\hat b$
measurement for a SM Higgs ($\hat a=1,\hat b=0$) with $\mhsm=110\gev$
for the four luminosity/bunch-merging options outlined in the text.
Here, $\delta=\tan^{-1}{\hat b\over \hat a}$ and $r=\sqrt{\hat a^2+\hat b^2}$.
For small $\delta$, ${\hat b\over\hat a}\sim \delta$.
\label{samplecasei}}
\end{center}\end{figure}

To gain a quantitative understanding of how successful such
a strategy for determining the CP-nature of the $\h$ can be,
let use define $(\hat a,\hat b)={(a,b)/\left(gm_\mu/2m_W\right)}$
and give contours at $\Delta\chi^2=1,4,6.635,9$ 
in the $\delta=\tan^{-1}{\hat b\over\hat a}$, $r=\sqrt{\hat a^2+\hat
  b^2}$ parameter space. We
define $I$ as the proton source intensity enhancement 
relative to the standard value implicit 
for the earlier-given benchmark luminosities. We compare four cases:
(i) the case of $P=0.2$, $L=0.15\fbi$, which
corresponds to $I=1$ and the 
polarization level naturally achieved without any special
selection against slow muons; (ii) 
we maintain the same proton intensity, $I=1$, select faster muons
to the extent that it becomes possible to 
merge neighboring muon bunches,  leading to $P^m(I=1)\sim 0.39$ and
$L=0.075\fbi$; (iii) we increase the proton 
source intensity by a factor of two, $I=2$, 
while selecting faster muons and
merging the bunches, corresponding to $P^m(I=2)\sim 0.48$ and
$L=0.075\fbi$; finally (iv) we employ $I=3$
and use so-called `just-full bunches', corresponding
to $P^f(I=3)\sim 0.45$ and $L=0.15\fbi$. 
Results in the case of a SM Higgs boson with $\mhsm=130\gev$
are presented in Fig.~\ref{samplecasei}. One sees
that a 30\% ($1\sigma$) measurement of $\hat b/\hat a$ is possible without
increased proton source intensity, using the simple technique
of selecting fast muons and performing bunch merging. An $\lsim 20\%$
measurement would require a moderately enhanced proton
source intensity.

After studying a number of cases, the overall conclusion of
\cite{Grzadkowski:2000hm} is that this procedure will provide
a good CP determination (superior to other techniques)
provided one merges bunches
and compensates for the loss of luminosity associated
with selecting only energetic muons 
(so as to achieve high average polarization) 
by having a proton source that is at least two
times as intense as that needed for the studies discussed
in previous sections (that do not require large transverse polarization).

\section{Conclusions and Planning for Future Facilities}

Around 2006 the LHC will begin taking data, hopefully revealing the path that 
particle physics will take in the next century. At the moment there are a 
few experimental hints suggesting that a Higgs boson might be just around 
the corner, and there are intriguing indications from the anomalous 
magnetic moment of the muon that supersymmetric particles may be easily 
detected at the LHC. This scenario would present a strong argument for the 
construction of a LC to study this interesting physics which would be 
at a scale light enough to be probed. 
A muon collider could play a crucial role in 
several ways. First, a $s$-channel light-Higgs factory would
provide crucial precision measurements of the $\hl$
properties, including the only accurate measurement of its $\mupmum$
coupling. Deviations of these properties with
respect to expectations for the SM Higgs boson can, 
in turn, impose critical constraints on the masses
of heavier Higgs bosons and other SUSY parameters.
Among other things, the heavier Higgs bosons might be
shown to definitely lie within reach of muon collider $s$-channel production.
Further, it could be that the heavier $\hh$ and $\ha$ 
cannot be detected at the LHC or LC (a scenario that arises, 
in the MSSM for example, for moderate
$\tanb$ values and $\mha\sim\mhh\gsim 250\gev$). Since
their detection in $s$-channel production at the muon collider would
be relatively certain, the muon collider would be an
essential component in elucidating the full physics of the Higgs sector.
Further, there are even (non-supersymmetric) scenarios in which 
one only sees a SM-like Higgs as the LHC and LC probe scales below a TeV,
but yet muon collider Higgs factory studies would reveal additional
Higgs bosons. Using $s$-channel Higgs production, 
a muon collider would also provide particularly
powerful possibilities for studying the CP nature of the
Higgs boson(s) that are found. Such CP determination might be absolutely
crucial to a full understanding of the Higgs sector.
 Finally, one should not forget that the muon collider
might prove to be the best approach to achieving
the highest energies possible in the least amount of 
time. Construction of a Higgs factory would be a vital link
in the path to high energy.


\begin{thebibliography}{99} 

\bibitem{Barate:2000ts}
R.~Barate {\it et al.}  [ALEPH Collaboration],
Phys.\ Lett.\ B {\bf 495}, 1 (2000)
[hep-ex/0011045].

\bibitem{Abreu:2001fw}
P.~Abreu {\it et al.}  [DELPHI Collaboration],
Phys.\ Lett.\ B {\bf 499}, 23 (2001)
[hep-ex/0102036].

\bibitem{Acciarri:2000ke}
M.~Acciarri {\it et al.}  [L3 Collaboration],
Phys.\ Lett.\ B {\bf 495}, 18 (2000)
[hep-ex/0011043].

\bibitem{Abbiendi:2001ac}
G.~Abbiendi {\it et al.}  [OPAL Collaboration],
Phys.\ Lett.\ B {\bf 499}, 38 (2001)
[hep-ex/0101014].\cite{Barate:2000ts,Abreu:2001fw,Acciarri:2000ke,Abbiendi:2001ac,Okpara:2001jf}

\bibitem{Okpara:2001jf}
A.~N.~Okpara,
hep-ph/0105151.

\bibitem{Brown:2001mg}
H.~N.~Brown {\it et al.}  [Muon g-2 Collaboration],
Phys.\ Rev.\ Lett.\  {\bf 86}, 2227 (2001)
[hep-ex/0102017].

\bibitem{Czarnecki:2001pv}
A.~Czarnecki and W.~J.~Marciano,
hep-ph/0102122.

\bibitem{Ankenbrandt:1999as}
C.~M.~Ankenbrandt {\it et al.},
Phys.\ Rev.\ ST Accel.\ Beams{\bf 2}, 081001 (1999)
[physics/9901022].


\bibitem{Raja:1998ip}
R.~Raja and A.~Tollestrup,
Phys.\ Rev.\ D {\bf 58}, 013005 (1998)
[hep-ex/9801004].

\bibitem{Barger:1997jm}
V.~Barger, M.~S.~Berger, J.~F.~Gunion and T.~Han,
Phys.\ Rept.\ {\bf 286}, 1 (1997)
[hep-ph/9602415].


\bibitem{Barger:1995hr}
V.~Barger, M.~S.~Berger, J.~F.~Gunion and T.~Han,
Phys.\ Rev.\ Lett.\ {\bf 75}, 1462 (1995)
[hep-ph/9504330].

\bibitem{Chankowski:2000an}
P.~Chankowski \emph{et al.},
{\sl Phys. Lett.} {\bf B496} (2000) 195 [hep-ph/0009271].

\bibitem{Autin:1999ci}
B.~Autin et.al.,
CERN-99-02.

\bibitem{Murray:2001es}
W.~J.~Murray,
hep-ph/0104268.

\bibitem{Taus} V. Barger, T. Han and C.G. Zhou, 
Phys. Lett. {\bf B480}, 140 (2000).

\bibitem{TauDecay} B. K. Bullock, K. Hagiwara and A. D. Martin, 
Nucl. Phys. {\bf B395}, 499 (1993).

\bibitem{ourstudy} D. Asner, J. Gronberg, J. Gunion and T. Hill,
in preparation. For a summary, see the $\gam\gam$ collider
section of the Snowmass 2001 Proceedings.

\bibitem{Battaglia:2000jb}
M.~Battaglia and K.~Desch,
hep-ph/0101165.

\bibitem{Gunion:1996cn}
J.~F.~Gunion, L.~Poggioli, R.~Van Kooten, C.~Kao and P.~Rowson,
hep-ph/9703330.


\bibitem{Groom:2000in}
D.~E.~Groom {\it et al.}  [Particle Data Group Collaboration],
Eur.\ Phys.\ J.\ C {\bf 15}, 1 (2000).

\bibitem{gunionmumu} 
J.~F.~Gunion,
``Detecting and studying Higgs bosons,'' in {\it Perspectives
on Higgs Physics, II}, ed. G.L. Kane, World Scientific Publishing (1997),
hep-ph/9705282.


\bibitem{Berger:2001et}
M.~S.~Berger,
hep-ph/0105128.

\bibitem{Berger:1990hg}
M.~S.~Berger,
Phys.\ Rev.\ D {\bf 41}, 225 (1990).


\bibitem{Gunion:1997cc}
J.F.~Gunion and J.~Kelly,
{\sl Phys. Rev.} {\bf D56} (1997) 1730
[hep-ph/9610495].

\bibitem{Gunion:1996qd}
J.F.~Gunion and J.~Kelly,
hep-ph/9610421.

\bibitem{Feng:1997xv}
J.L.~Feng and T.~Moroi,
{\sl Phys. Rev.} {\bf D56} (1997) 5962
[hep-ph/9612333].



\bibitem{Barger:1997pv}
V.~Barger, M.~S.~Berger, J.~F.~Gunion and T.~Han,
Phys.\ Rev.\ Lett.\ {\bf 78}, 3991 (1997)
[hep-ph/9612279].

\bibitem{Miller:2001bi}
D.~J.~Miller, S.~Y.~Choi, B.~Eberle, M.~M.~Muhlleitner and P.~M.~Zerwas,
Phys.\ Lett.\ B {\bf 505}, 149 (2001)
[hep-ph/0102023].

\bibitem{farrisgunion} T. Farris and J.F. Gunion, in preparation.

\bibitem{gunionucla} J.F. Gunion, in
``Transparency Book'', Higgs Factory Workshop, Feb. 28 -- March 1, 2001,
UCLA.


\bibitem{Grzadkowski:1995rx}
B.~Grzadkowski and J.~F.~Gunion,
Phys.\ Lett.\ B {\bf 350}, 218 (1995)
[hep-ph/9501339].

\bibitem{Asakawa:2001es}
E.~Asakawa, S.~Y.~Choi and J.~S.~Lee,
Phys.\ Rev.\ D {\bf 63}, 015012 (2001)
[hep-ph/0005118].

\bibitem{Pilaftsis:1998dd}
A.~Pilaftsis,
Phys.\ Lett.\ B {\bf 435}, 88 (1998)
[hep-ph/9805373].

\bibitem{Pilaftsis:1998pe}
A.~Pilaftsis,
Phys.\ Rev.\ D {\bf 58}, 096010 (1998)
[hep-ph/9803297].

\bibitem{Grzadkowski:2000hm}
B.~Grzadkowski, J.~F.~Gunion and J.~Pliszka,
Nucl.\ Phys.\ B {\bf 583}, 49 (2000)
[hep-ph/0003091].

\end{thebibliography}
\end{document}